\documentclass[twocolumn]{aastex62}
\submitjournal{AJ}

\shorttitle{NGC~1533, IC~2038 and IC~2039: an interacting triplet in the Dorado group
}
\shortauthors{Cattapan et al.}

\begin{document}

\title{
VEGAS: a VST Early-type GAlaxy Survey. IV. \\ 
NGC~1533, IC~2038 and IC~2039: an interacting triplet in the Dorado group
}

\correspondingauthor{Arianna Cattapan}
\email{cattapan.arianna@gmail.com}

\author[0000-0002-0786-7307]{Arianna Cattapan}
\affil{INAF-Astronomical Observatory of Capodimonte, Salita Moiariello 16, 80131, Naples, Italy}

\author{Marilena Spavone}
\affil{INAF-Astronomical Observatory of Capodimonte, Salita Moiariello 16, 80131, Naples, Italy}

\author{Enrichetta Iodice}
\affil{INAF-Astronomical Observatory of Capodimonte, Salita Moiariello 16, 80131, Naples, Italy}

\author{Roberto Rampazzo}
\affil{INAF-Astronomical Observatory of Padova, Vicolo dell’Osservatorio 8, I-36012, Asiago, Italy}

\author{Stefano Ciroi}
\affil{Dept. of Physics and Astronomy ``G. Galilei"- University of Padova, Vicolo dell'Osservatorio 3, I-35122, Padova, Italy}

\author{Emma Ryan-Weber}
\affil{Centre for Astrophysics and Supercomputing, Swinburne University of Technology, Hawthorn, Victoria 3122, Australia}

\author{Pietro Schipani}
\affil{INAF-Astronomical Observatory of Capodimonte, Salita Moiariello 16, 80131, Naples, Italy}

\author{Massimo Capaccioli}
\affil{University of Naples Federico II, C.U. Monte Sant’Angelo, via Cinthia, 80126, Naples, Italy}

\author{Aniello Grado}
\affil{INAF-Astronomical Observatory of Capodimonte, Salita Moiariello 16, 80131, Naples, Italy}

\author{Luca Limatola}
\affil{INAF-Astronomical Observatory of Capodimonte, Salita Moiariello 16, 80131, Naples, Italy}

\author{Paola Mazzei}
\affil{INAF-Astronomical Observatory of Padova, Vicolo dell’Osservatorio 5, I-35122, Padova, Italy}

\author{Enrico V. Held}
\affil{INAF-Astronomical Observatory of Padova, Vicolo dell’Osservatorio 8, I-36012, Asiago, Italy}

\author{Antonietta Marino}
\affil{INAF-Astronomical Observatory of Padova, Vicolo dell’Osservatorio 5, I-35122, Padova, Italy}






\begin{abstract}
This paper focuses on NGC~1533 and the pair IC~2038 and IC~2039 in Dorado a nearby, clumpy, still un-virialized group.
We obtained their surface photometry from deep OmegaCAM@ESO-VST images in {\it g} and {\it r}  bands.
For NGC~1533, we map the surface brightness down to $\mu_g \simeq 30.11$~mag~arcsec$^{-2}$ and $\mu_r \simeq 28.87$~mag~arcsec$^{-2}$ and out to about $4R_e$.
At such faint levels the structure of NGC~1533 appear amazingly disturbed with clear structural asymmetry between inner and outer isophotes in the North-East direction. We detect new spiral arm-like tails in the outskirts, which might likely be the signature of a past interaction/merging event. 
Similarly, IC~2038 and IC~2039 show tails and distortions indicative of their ongoing interaction.  
Taking advantages of deep images, we are able to detect the optical counterpart to the HI gas. 
The analysis of the new deep data suggests that NGC~1533 had a complex history made of several interactions with low-mass satellites that generated the star-forming spiral-like structure in the inner regions and are shaping the stellar envelope. 
In addition, the VST observations show that also the two less luminous galaxies, IC~2038 and IC~2039, are probably interacting each-other and, in the past, IC~2038 could have also interacted with NGC~1533, which stripped away gas and stars from its outskirts.
The new picture emerging from this study is of an interacting triplet, where the brightest galaxy NGC~1533 has ongoing mass assembly in the outskirts.

\end{abstract}



\section{Introduction} \label{sec:intro}

Early-type galaxies (ETGs=Es+S0s) populating dense clusters represent the end-product of a long evolution. Most of them have exhausted their gas reservoir so that  star formation quenched long time ago. These galaxies represent the {\it template} of the passively evolving ETGs \citep[see e.g.][for Virgo members]{Bressan2006}, the so-called ``red and dead'' galaxy population \citep[see e.g.][and references therein]{Boselli2014}. How and when the processes leading to the final quenching of star formation have happened is still an open and widely debated question. Indeed, supporters of either a {\it gentle} or a {\it sudden} quenching process are found in the literature \citep[see e.g.][]{Eales2017,Weigel2017,Oemler2017}.

In less dense environments, a series of observations at different wavelengths have by now shown that ETGs are, in general, ``still alive'' or at least ``undead'' with respect to their cluster counterparts.  At optical wavelengths, \citet[][and references therein]{Clemens2006} analyzing  Lick line-strength indices in central regions found that ETGs in the field are younger than those in environments typical of clusters. Young ETGs are those in which  a recent episode of star formation occurred, {\it rejuvenating} their stellar population. Refining their  analysis on a larger sample, \citet{Clemens2009} found that field ETGs are younger than their cluster counterparts by $\sim2$~Gyrs. At mid-infrared wavelengths {\it Spitzer}-{\tt IRS} \citep{Werner2004,Houck2004} observations revealed the presence of Polycyclic Aromatic Hydrocarbons (PAHs) in several nuclear region of ETGs, in particular S0s \citep[see e.g.][and references therein]{Kaneda2005,Panuzzo2011,Rampazzo2013}. The PAHs  witness a recent star formation activity.  They arise from fresh carbonaceous material that is continuously released by a population of carbon stars, formed in a rejuvenation episode that occurred within the last few Gyrs \citep{Vega2010}. The {\it Galaxy Evolution Explorer} ({\tt GALEX}) \citep{Morrissey2007} via UV surveys as well as through deeper single target observations has contributed to many breakthroughs in revealing ongoing star formation in ETGs \citep{Rampazzo2007,Jeong2009,Salim2010,Marino2011a,Marino2011b}. {\tt GALEX} showed that star formation is not confined to the nuclear part but extends to the galaxy outskirts \citep[e.g. in XUV disks][]{Thilker2008}.

In this context, the study of the population of ETGs member of nearby, poor environments, like groups,  is of overwhelming importance. In groups, we attend to  both rejuvenation and quenching phenomena, that are found only  in the outskirts of clusters. Although in groups these mechanisms could differ from those acting in clusters \citep{Boselli2014}, because of the lower velocity dispersion of member galaxies, the importance of evolutionary mechanisms in groups is twofold. Groups contain more than $60\%$ of the galaxies in the nearby Universe and forge ETGs that will be accreted by clusters. 

The history of galaxy transformation from active into passively evolving systems is written in the color magnitude diagrams (CMd hereafter) of their galaxy populations. Particularly evident in optical vs. UV CMd is the area inhabited by ETGs  i.e. the red sequence, although they could be found also in the so-called {\it green valley}. The red sequence represents the final destination of galaxies which, during their evolution, have passed from the blue cloud, inhabited by star-forming galaxies, via the green valley \citep{Mazzei2014a,Mazzei2014b,Mazzei2018a,Mazzei2018b}. CMds tell us also the evolutionary phase of the group. Young groups have a significant blue cloud at the expense of a depopulated red sequence. In evolved groups the red sequence is well-defined basically at all magnitudes of the galaxy population \citep{Marino2016}.  

In the recent years, a big effort was made to develop deep photometric surveys aimed at studying galaxy structures out to the 
regions of the stellar halos \citep[e.g.][]{Ferrarese2012,vDokkum2014,Duc2015,Munoz2015,Merritt2016,Trujillo2016,Mihos2017}. 
The VST Early-type GAlaxy Survey\footnote{See http://www.na.astro.it/vegas/VEGAS/Welcome.html} \citep[VEGAS,][]{Capaccioli2015} has occupied in the last years a pivotal role in the field, by providing new insight on the faint regions of galaxies in all environments. 
VEGAS is a multiband {\it u}, {\it g}, {\it r} and {\it i} imaging survey, able to map the surface brightness of galaxies down to the azimuthally averaged surface brightness $\mu_g\sim30$~mag~arcsec$^{-2}$ and out to $\sim10 R_e$. 
Therefore, at these faint levels in surface brightness, with the VEGAS data we are able to study the faintest regions of the galaxy outskirts and to address the build up history of the stellar halo by comparing the surface brightness profiles and the stellar mass fraction with the predictions of cosmological galaxy formation theories \citep{Iodice2016,Spavone2017b,Iodice2017a,Iodice2017b,Spavone2018}. 
Moreover, thanks to the coverage of different morphological types, masses and environments in VEGAS, we have started a study dedicated to the synoptic analysis of globular clusters (GCs) in different host galaxies \citep{Dabrusco2016,Cantiello2018}.
The  VEGAS sample is made up by 58 ETGs brighter than M$_B = -21$~mag, in different environments, including giant cD galaxies in the core of clusters, ETGs in group of galaxies and  in low-density regions, in the local volume within $D\le 54$~Mpc/h. 
In this paper we present the new deep data for one of the VEGAS target: the Dorado group of galaxies.

\subsection{The Dorado Group}
\begin{figure}
	\gridline{\fig{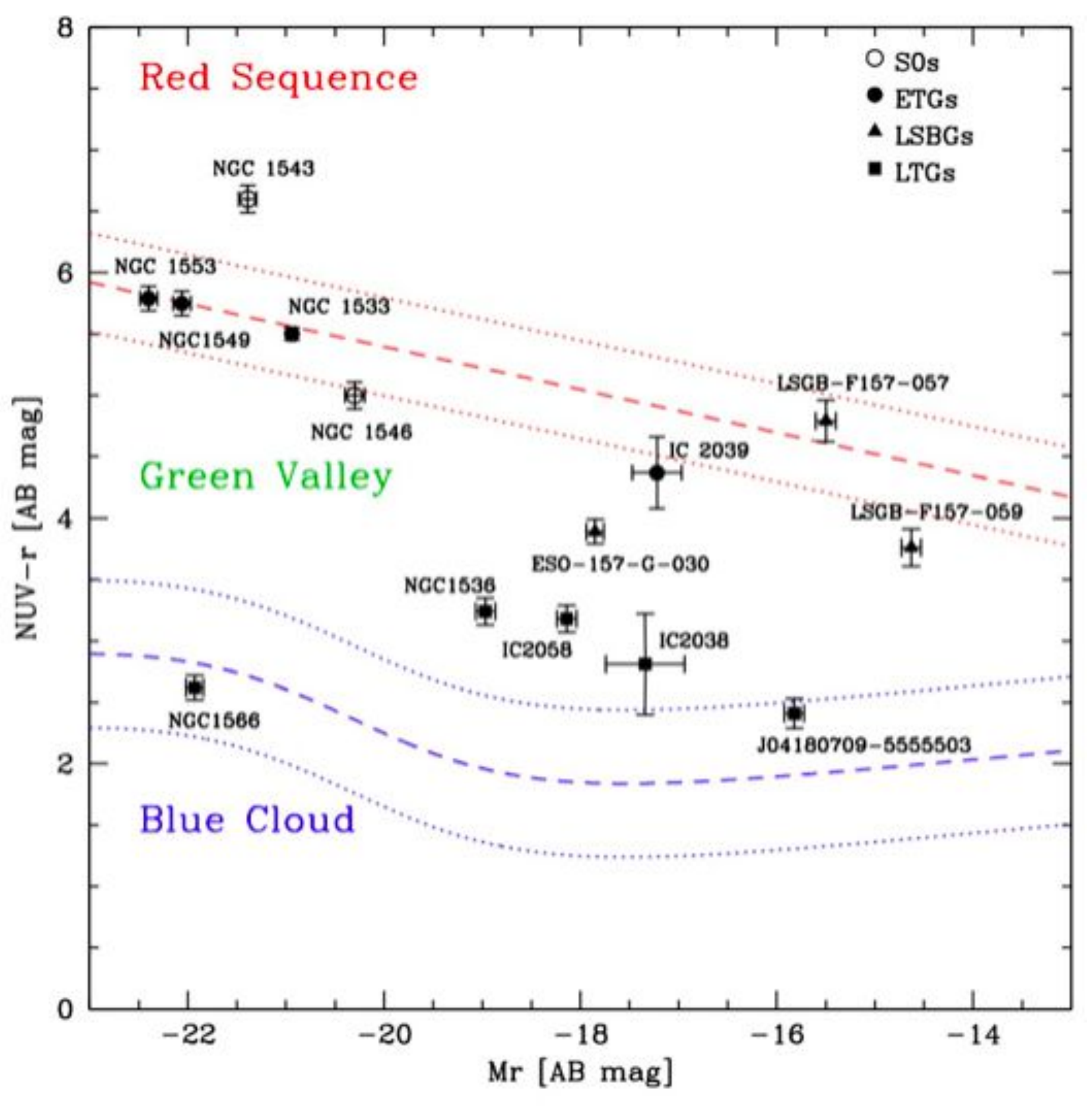}{0.475\textwidth}{}}
	\caption{M$_r$ versus (NUV-{\it r}) Color Magnitude diagram. The \citet{Wyder2007} fits (dashed lines), plus the error estimation (dotted lines), to the Red sequence and Blue cloud are shown. Empty circles mark lenticulars (S0s), full circles mark early-type galaxies (ETGs), full triangles mark low surface brightness galaxies (LSBGs) and full squares mark late-type galaxies (LTGs). \label{fig:cmd}}
\end{figure}

The Dorado group, firstly identified by \citet{deVaucouleurs1975}, has a backbone members structure which has been provided, more recently, by \citet{Firth2006}.  In that study Dorado results to be a loose and clumpy group,  un-virialized, centered at about $\alpha=4^h$~$17^m$~$01.8^s$ and $\delta=-55^{d}$~$42^{m}$~$46^{s}$~(J2000), extending for several degrees ($\approx 10^{\circ}\times10^{\circ}$) in projection. The distance of the Dorado group has a large uncertainty due also to the membership definition: from $16.9$~Mpc (V$_{hel}=1246\pm39$~km~s$^{-1}$) given by \citet{Firth2006} to $21$~Mpc by \citet{Brough2006}. Following  \citet{Firth2006} the velocity dispersion of the member galaxies ($20$ in total) is $222\pm17$~km~s$^{-1}$ with a crossing time of $0.126\pm0.006$~H$^{-1}_0$.
The group CMd, shown in Figure~\ref{fig:cmd}, has a well-defined red sequence, a nearly depopulated blue cloud and a set of member galaxies crossing  the green valley. The brightest galaxies of the group are NGC~1549 (E$0-1$ {\tt NED\footnote{{\tt NED} Nasa/IPAC Extragalactic Database https://ned.ipac.caltech.edu}}) and NGC~1553 (SA0$^0$(r) {\tt NED}), both showing a wide system of shells \citep{Malin1983}.  There are many galaxies showing interaction signatures suggesting that the group is going through a strongly evolving phase suggested also by the M$_r$ vs. (NUV-{\it r}) CMd \citep{Rampazzo2018b,Cattapan2018} described above.  The group has a significant candidate dwarf galaxy population concentrated towards the group center detected by \citet{Carrasco2001}. The dwarf galaxies' population would provide direct indication of the collapsed halo and the ratio giant vs. dwarfs is larger when the group is dominated by ETGs \citep{Tully2015a}, as would be in the case of Dorado.\\

The subjects of this paper are NGC~1533, the fourth bright galaxy member of Dorado, and the two nearby fainter interacting galaxies IC~2038  and IC~2039, the pair AM~0407-560 \citep{Arp1982}.  The average radial velocity of the three galaxies is $771\pm54$~km~s$^{-1}$ (see Table~\ref{tab:mw-prop}) indicating that they belong to the same substructure in the group. In addition, the three galaxies are embedded in a cloud of HI identified with the HIPASS name J$0409-56$ \citep{RyanWeber2003b} whose mass is $0.933\times10^9$~$\mathcal{M}_{\odot}$ (using a distance of D$=7.6$~Mpc) \citep{RyanWeber2003b}. The shape and the velocity field of the HI cloud \citep{RyanWeber2003a} support the view of a co-evolution of NGC~1533 and IC~2039 \citep{RyanWeber2004,Bekki2005,Werk2010}. 
The morphology of the lenticular NGC~1533 is complex as described by the \citet{Comeron2014} classification: it is a (RL)SB0$^\circ$ since it shows a complete, regular ring, a lens and a bar, seen nearly face on. Both {\tt GALEX} \citep{Marino2011b} and {\it Swift}-{\tt UVOT} observations \citep{Rampazzo2017} have shown that NGC~1533 hosts ongoing star formation.

\begin{deluxetable*}{lccccccccc}
	\tablenum{1}
	\tablecaption{Basic properties of galaxy sample\label{tab:mw-prop}}
	\tablewidth{0pt}
	\tablehead{
		\colhead{Galaxy} & 
		\colhead{RA} & 
		\colhead{Decl.} & 
		\colhead{Morph.} &
		\colhead{D} &
		\colhead{B$_T$} & 
		\colhead{V$_{hel}$} &
		\colhead{$\sigma$} & 
		\colhead{P.A.} & 
		\colhead{$\epsilon$}\\
		\colhead{} & 
		\colhead{[h~m~s]} & 
		\colhead{[$^{\circ}$~$^{\prime}$~$^{\prime\prime} $]} & 
		\colhead{Type} &
		\colhead{[mag]}&
		\colhead{[mag]} & 
		\colhead{[km~s$^{-1}$]} & 
		\colhead{[km~s$^{-1} $]} & 
		\colhead{[deg NE]} &
		\colhead{}\\
		\colhead{(1)} &
		\colhead{(2)} &
		\colhead{(3)} &
		\colhead{(4)} &
		\colhead{(5)} &
		\colhead{(6)} &
		\colhead{(7)} &
		\colhead{(8)} &
		\colhead{(9)} &
		\colhead{(10)}
	}
	\startdata
	NGC~1533 & $04$~$09$~$51.8$ & $-56$~$07$~$06$ & SB0$^-$ & $31.56\pm0.14$ & $11.79\pm0.15$ & $785\pm8$ & $177.5\pm4.5$ & $148.9$ & $0.37$\\
	IC~2038 & $04$~$08$~$53.7$ & $-55$~$59$~$22$ & Sd pec & $31.42$& $15.42\pm0.25$ & $712\pm52$ & \dots &$155.2$ & $0.74$\\
	IC~2039 & $04$~$09$~$02.4$ & $-56$~$00$~$42$ & S0$^0$: pec & $31.56$ &$14.99\pm0.14$ & $817\pm45$ & \dots & $124.3$ & $0.15$\\
	\enddata
	\tablecomments{Columns (2) and (3) are the J2000 coordinates; (4) morphological types 
		are from {\tt RC3}\footnote{Third Reference Catalogue of Bright Galaxies}; (5) the distance modulus from \citet{Tully2013} for NGC~1533, from \citet{Karachentsev2013} for IC~2038, and for IC~2039 we adopt the distance modulus of NGC~1533; (6) total apparent magnitude from {\tt HyperLeda}\footnote{http://leda.univ-lyon1.fr};  (7) radial heliocentric velocity 6df DR2 \citep{Firth2006};  (8) average value of central velocity dispersion from {\tt HyperLeda}; mean position angle measured North-East (9) and ellipticity (10) of isophotes measured at $\mu_B=25$~mag~arcsec$^{-2}$  {\tt HyperLeda}.}
\end{deluxetable*}

The paper is addressed to characterize, from accurate surface photometry, the structure of these galaxies up to the faintest surface brightness levels and to infer their possible interaction scenario. To this purpose we have obtained deep wide field exposures centered on NGC~1533 using the OmegaCAM@ESO-VLT in {\it g} and {\it r} SDSS filters.

The plan of the paper is the following.  Section~\ref{sec:O-R} reports about observations and reduction technique adopted. In Section~\ref{sec:SB} we present the performed isophotal analysis and in Section~\ref{sec:results} are reported the results obtained for each galaxy. In Section~\ref{sec:discussion} we discuss results in the light of the current literature. The conclusions are drawn in Section~\ref{sec:conc}.

\section{Observations and Data Reduction} \label{sec:O-R}

The data we present in this work are part of the VEGAS Survey, which is based on the ESO VLT Survey Telescope (VST) Guaranteed Time Observation (GTO) assigned at the Italian National Institute for Astrophysics (INAF). 
Our data-set was obtained with VST+OmegaCAM in visitor mode (run ID 0100.B-0168(A)), in dark time, as summarized in Table~\ref{tab:obs-log}.
All observations were obtained by using the standard diagonal dithering for the scientific frame, in order to cover the gaps of the camera.
Each science frame has an integration time of $300$~sec and we have obtained a total integration time of $2.2$~hrs in the {\it g} band and $1.3$~hrs in the {\it r}, in an arcsec-level seeing conditions (see Table~\ref{tab:obs-log}).
Data were processed with the {\it VST-Tube pipeline}, described in details by \citet{Grado2012} and \citet[Appendix A]{Capaccioli2015}.

\begin{deluxetable*}{ccccccc}
	\tablecaption{Observation log for {\it g} and {\it r} data \label{tab:obs-log}}
	\tablecolumns{7}
	\tablenum{2}
	\tablewidth{0pt}
	\tablehead{
		\colhead{Filter} 
		&\colhead{Date} 
		& \colhead{RA (J2000)}
		& \colhead{Decl. (J2000)}
		& \colhead{T$_{exp}$} 
		& \colhead{ Number combined}
		& \colhead{Seeing (FWHM)}\\
		\colhead{[SDSS]} 
		& \colhead{[YYYY-mm-dd]} 
		& \colhead{[deg]}
		& \colhead{[deg]}
		& \colhead{[sec]}
		& \colhead{frames}
		& \colhead{[arcsec]} \\
		\colhead{(1)} 
		& \colhead{(2)}
		& \colhead{(3)}
		& \colhead{(4)} 
		& \colhead{(5)} 
		& \colhead{(6)}
		& \colhead{(7)}
	}
	\startdata
	{\it g} & $2017-10-20/27$ & $62.444$ & $-56.101$ &  $7800$ & $26$ & $0.7893$ \\
	{\it r}  & $2017-10-20/27$ & $62.289$ & $-56.114$ & $4800$ & $16$ & $0.7852$ \\
	\enddata
	\tablecomments{(1) Filters in the SDSS-band; (2) dates of observations; (3) right ascension and declination (4) of the center of the field of view;  (5) total exposure time; (6) number of combined dithered frames; (7) median value of the seeing of the combined frames.}
\end{deluxetable*}

\begin{figure*}
	\gridline{\fig{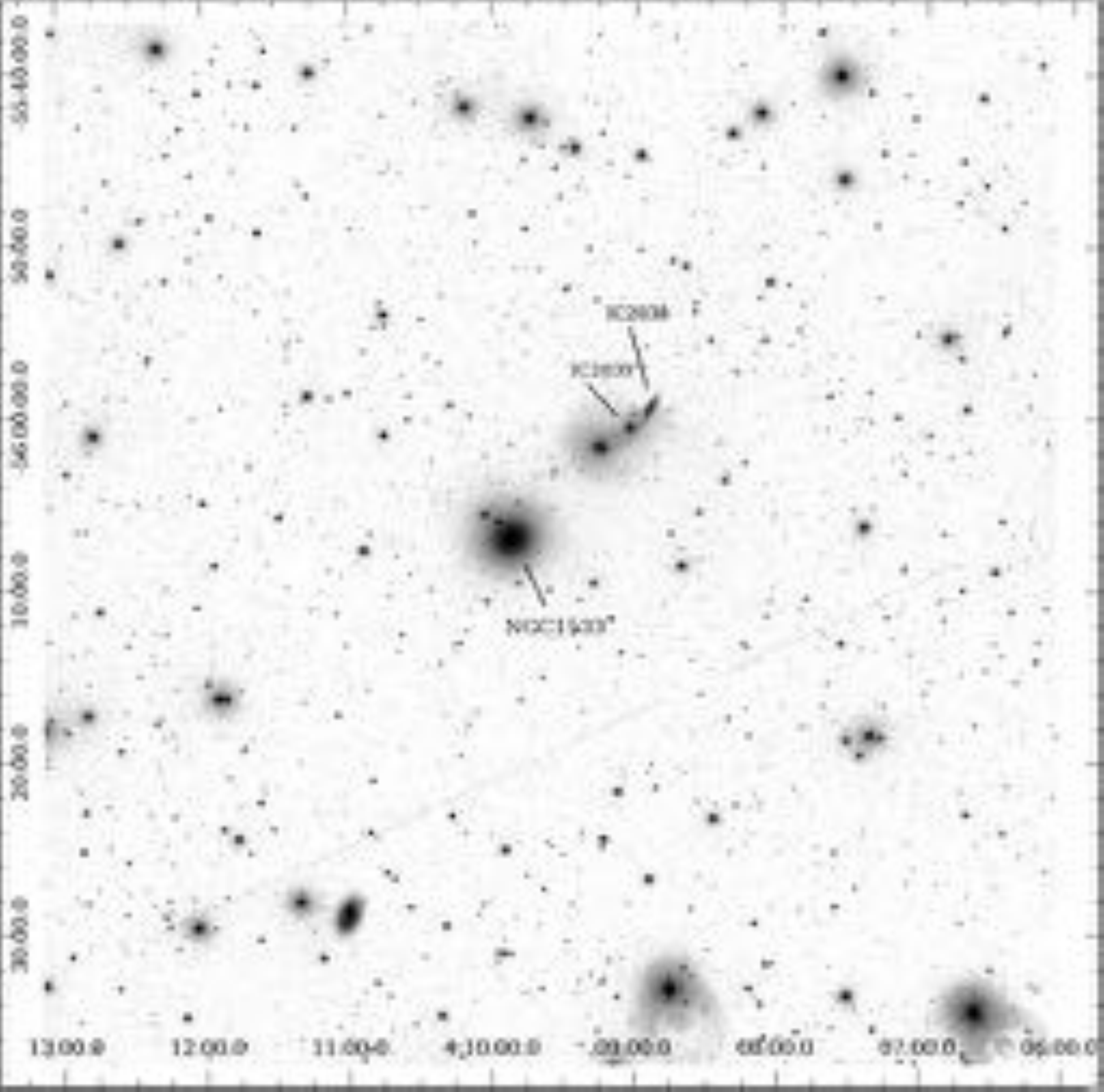}{0.405\textwidth}{}
		\fig{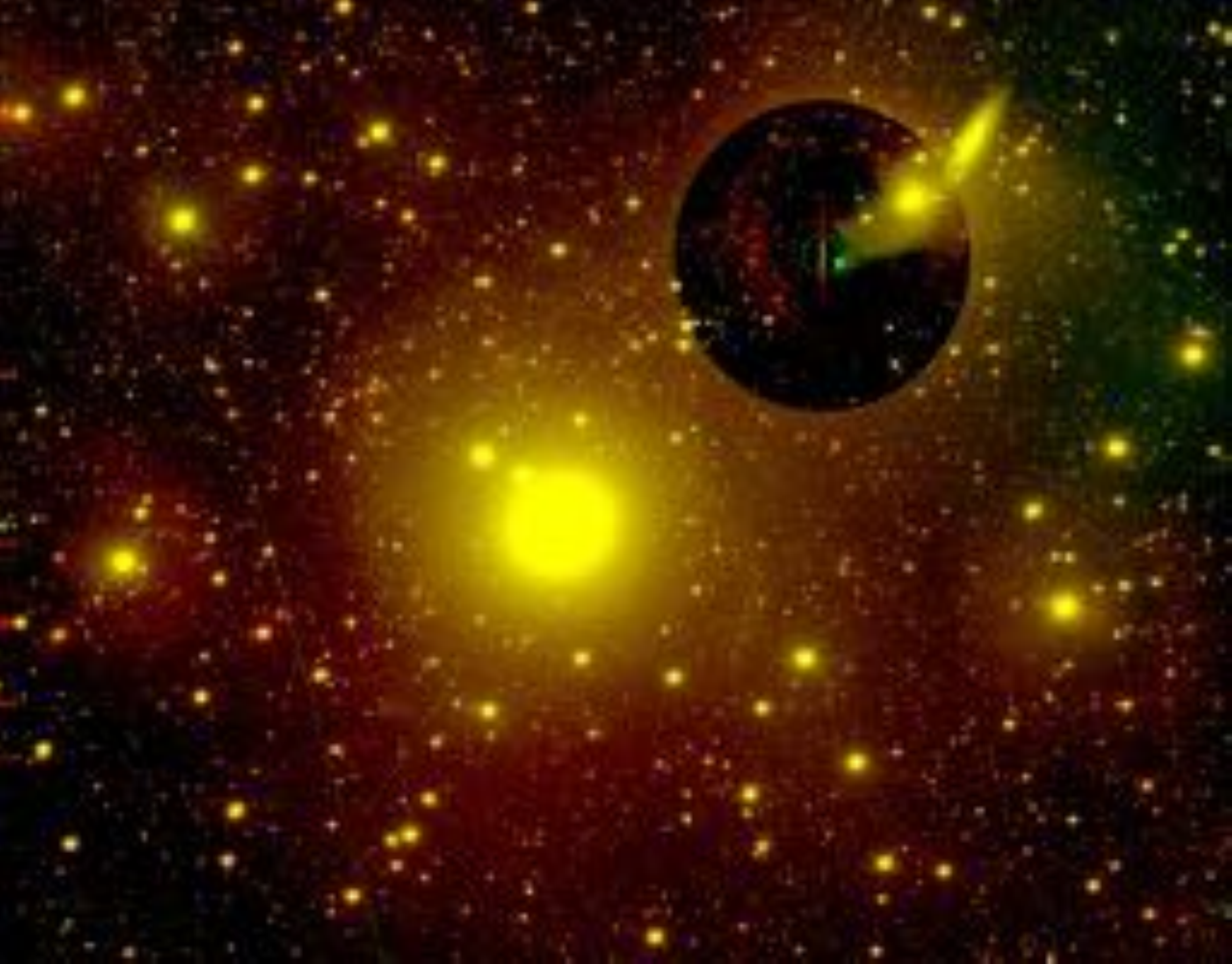}{0.51\textwidth}{}
	}
	\caption{{\it Left panel}: OmegaCAM@VST {\it g} band observed field of view, centered on NGC~1533. The image size is $\sim1^{\circ} \times1^{\circ}$. {\it Right panel}: color composite image, $\sim24^{\prime}\times19^{\prime}$, of NGC~1533, IC~2038 and IC~2039. North is at the top, east is on the left.  \label{fig:vstcolcomposite}}
\end{figure*}

Figure~\ref{fig:vstcolcomposite} shows the observed field of view in {\it g} band, centered on NGC~1533, and the color composite image using {\it g} (green channel) and {\it r} (red channel) band images from OmegaCAM@VST.

%

\subsection{Sky background subtraction}\label{subsec:SB}

For the final calibrated images obtained for the field around the brightest group member, NGC~1533, in each band, we have estimated and subtracted the sky background by using the IRAF\footnote{Image Reduction and Analysis Facility\\ http://iraf.noao.edu} task \textsc{imsurfit}. 
It  allows to fit the background contribution to the scientific images with a polynomial function.
Thanks to the large field of view of OmegaCAM@VST, we have estimated the background value by exploiting the region around NGC~1533, IC~2038 and IC~2039, where the contribution to the light of galaxies and background/foreground stars were masked.
The best fit of the background contribution was achieved using a third order Chebyshev polynomial function.

\begin{figure}
	\plotone{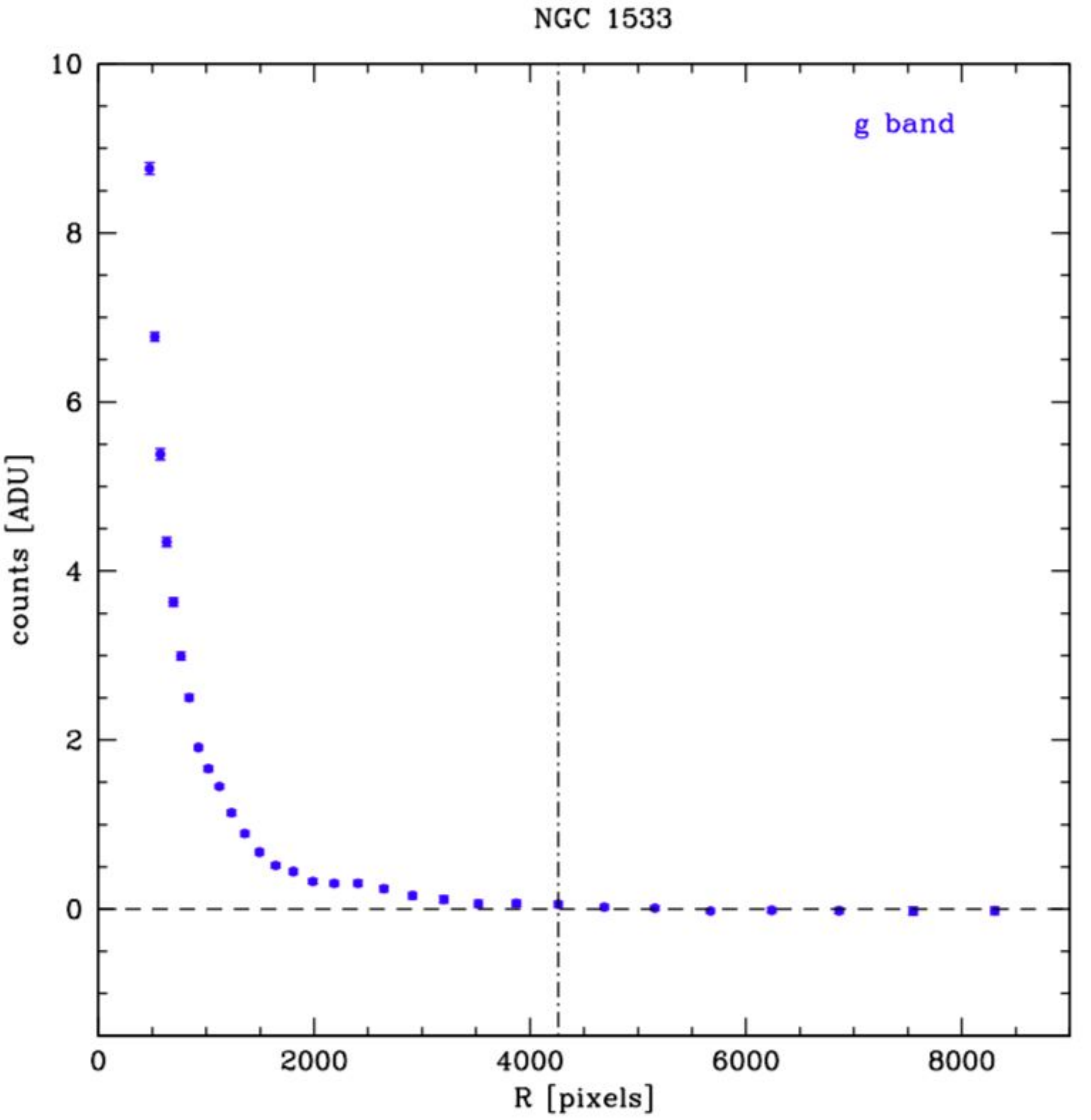}
	\caption{Azimuthally averaged intensity profile in counts as a function of the semi-major axis of \textsc{ellipse}, for {\it g} images of NGC~1533. The vertical dashed-dotted line indicates the outermost radius; the horizontal dashed line the mean residual background fluctuations.\label{fig:resback}}
\end{figure}

In the background-subtracted science frame, ``residual fluctuations'' of a few percent in flux could remain, due to the flux variation of the background during the night. These are estimated on the final stacked image and set the accuracy of the background-subtraction step. 
To this aim, we extracted the azimuthally averaged intensity profile centered on NGC~1533 with IRAF task \textsc{ellipse},
in each band, and we estimated the outermost radius $R_{lim}$, distance from the galaxy's center where the galaxy's light blends into the average residual background level. This is the residual level, which persists after subtracting the background surface obtained by the polynomial fit, and it is very close to zero. 
We estimated the average sky fluctuation levels ($\Delta_{sky}$) and the ``amplitudes'' of the distribution, i.e. the RMS of the mean value at one sigma. 
We assume as the latest valuable science point in the surface brightness profile the point where flux $\geq \Delta_{sky}$.
For NGC~1533, $R_{lim} \simeq 14\farcm91$, in the {\it g} band, while in {\it r} band $R_{lim} \simeq 13\farcm56$.
Figure \ref{fig:resback} shows an example of the above estimations in the {\it g} band.
This method, which is well tested in the literature and also on the previous VEGAS data and analysis 
\citep[see e.g.][]{Pohlen2006,Iodice2014,Capaccioli2015,Iodice2016,Iodice2017a,Spavone2017b,Spavone2018,Iodice2018}, 
allows to measure any residual fluctuations in the background-subtracted image and then take them into account in the luminosity and surface brightness measurements and related errors.

\section{Surface photometry: isophotal analysis and light distribution}\label{sec:SB} 

\begin{figure*}
	\gridline{\fig{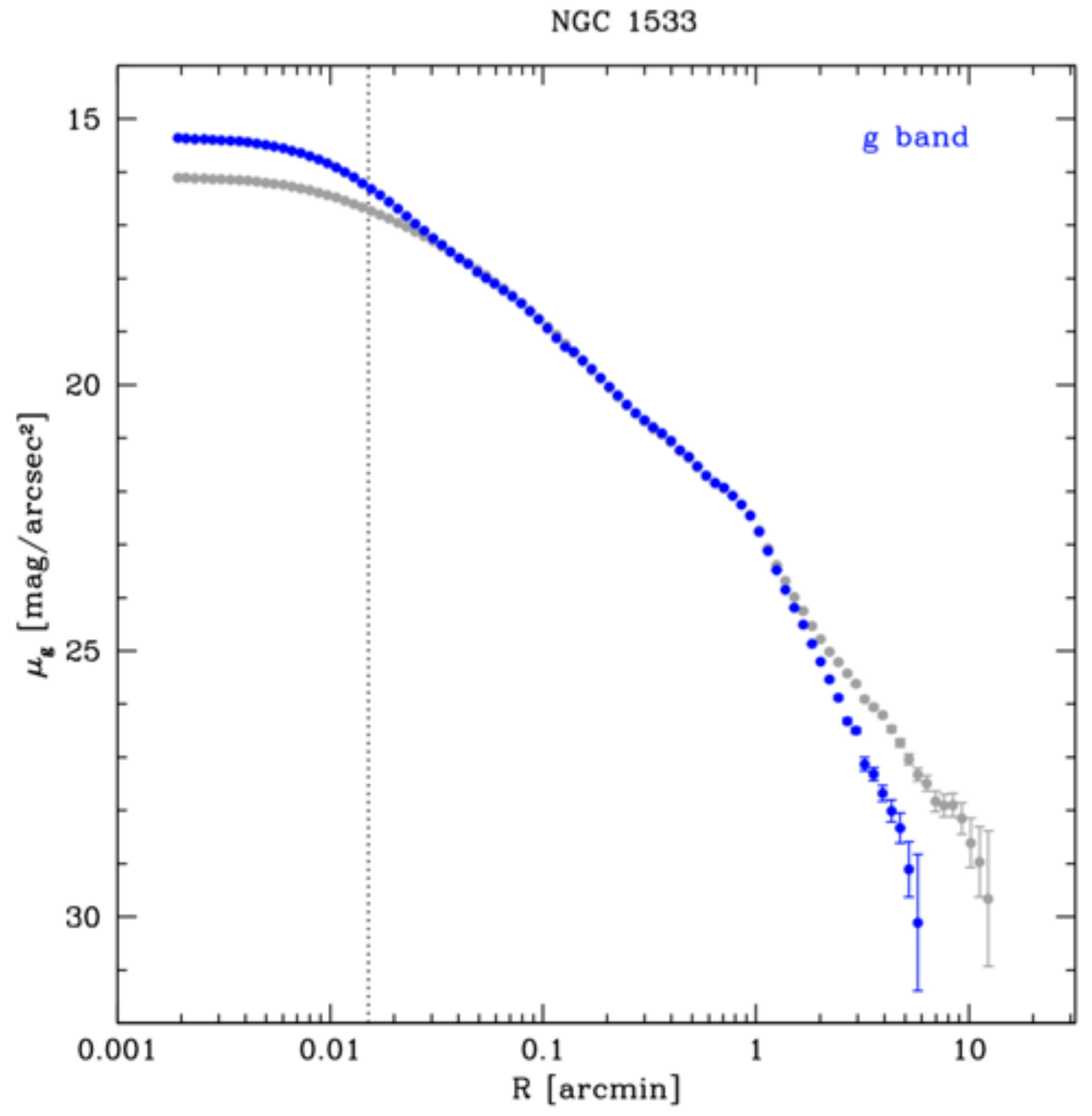}{0.43\textwidth}{}
	\fig{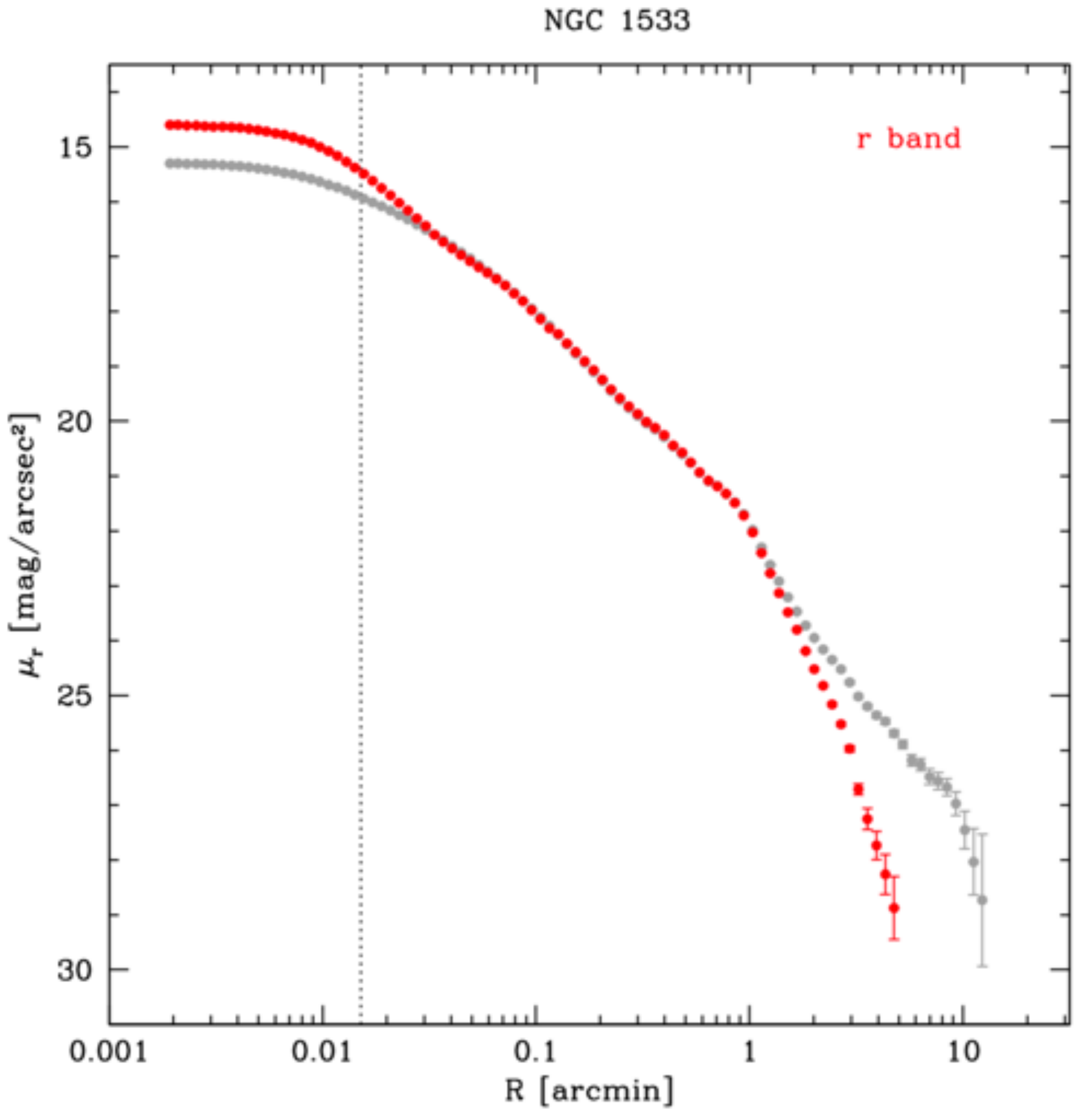}{0.43\textwidth}{}
	}
	\caption{{\it g} band ({\it left panel}) and {\it r} band ({\it right panel}) azimuthally averaged surface brightness profiles as a function of the logarithmic isophote semi-major axis. The gray dots correspond to the original profiles, the blue ({\it g} band) and red ({\it r} band) dots indicate the PSF deconvolved profiles.}\label{fig:1533psf}
\end{figure*}

The isophotal analysis is performed with \textsc{ellipse} task in IRAF \citep{Jedrzejewski1987} up to the limiting radius on the final stacked images in each band.  
\textsc{ellipse} provides both the geometrical parameters and the azimuthally averaged light distribution profiles. The error on the surface brightness accounts for the errors on the residual background (Subsection~\ref{subsec:SB}) and the zero point errors $ZP_g =24.5036 \pm 0.0018$ and $ZP_r = 24.3820 \pm 0.0012$ \citep{Capaccioli2015}. 
For each galaxy, i.e. for NGC~1533 and for the two companion galaxies IC~2038 and IC~2039, we have derived the azimuthally averaged surface brightness profiles and the shape parameters of the isophotes: the ellipticity ($\epsilon$), the position angle (P.A.) and the $a_4$ and $b_4$ \citep[see][]{Jedrzejewski1987}.
In the outer parts of galaxies, for NGC~1533 at radii larger than $\sim 5 \arcmin$, the signal-to-noise ratio is very low because the galaxy light begins to become comparable to the background residual fluctuations, therefore all the geometrical parameters are affected by strong fluctuations and large errors.

The results are described in Section~\ref{sec:results} and a detailed analysis based on the geometrical parameters is reported in Appendix~\ref{apx:A}.
The total magnitudes and the effective radii are estimated from the galaxy growth curve. The absolute magnitudes and the total luminosity in solar units\footnote{Absolute magnitudes of the Sun are reported in the following Website: http://mips.as.arizona.edu/~cnaw/sun.html} come from previous data and considering the galaxies distances (see Table~\ref{tab:mw-prop}). For the three galaxies these parameters are listed in Table~\ref{tab:results1}.

	\begin{deluxetable*}{ccccccccccc}
		\tabletypesize{\footnotesize}
		\tablecaption{Main results for NGC~1533, IC~2038 and IC~2039} \label{tab:results1}
		\tablecolumns{6}
		\tablenum{3}
		\tablewidth{0pt}
		\tablehead{
			\colhead{Galaxy} 
			& \colhead{m$_{tot,g}$} 
			& \colhead{m$_{tot,r}$}
			& \colhead{R$_{e,g}$} 
			& \colhead{R$_{e,r}$} 
			& \colhead{M$_g$}
			& \colhead{M$_r$}
			& \colhead{L$_g$}
			& \colhead{L$_r$}\\
			\colhead{} 
			&\colhead{[mag]} 
			& \colhead{[mag]} 
			& \colhead{[arcsec]} 
			& \colhead{[arcsec]}
			&\colhead{[mag]} 
			&\colhead{[mag]} 
			&\colhead{$\left[ L_{\odot} \right]$}
			&\colhead{$\left[ L_{\odot} \right]$}\\
			\colhead{(1)} 
			&\colhead{(2)}
			& \colhead{(3)}
			& \colhead{(4)}
			& \colhead{(5)}
			& \colhead{(6)}
			& \colhead{(7)}
			& \colhead{(8)}
			& \colhead{(9)}
		}
		\startdata
		NGC~1533&$10.48\pm0.05$ & $9.60\pm0.05$ & $64.17\pm0.16$ & $78.20\pm0.41$  & $-21.08 \pm 0.34$ & $-21.96 \pm 0.34$ & $2.99\times10^{10}$ & $ 4.40\times10^{10}$ \\
		IC~2038&$14.29\pm0.08$ & $13.94\pm0.05$ & \dots & \dots & $-17.13\pm0.08$ & $-17.48\pm0.05$ & $7.87\times10^8$ &$7.11\times10^8$ \\
		IC~2039&$14.41\pm0.05$ & $13.84\pm0.05$ & \dots & \dots & $-17.15\pm0.05$ & $-17.72\pm0.05$ & $8.01\times10^8$ & $8.86\times10^8$\\
		\enddata
		\tablecomments{(1) Galaxy name. Total magnitude in {\it g}, (2), and {\it r}, (3), bands. Effective radius in {\it g}, (4), and {\it r}, (5), bands. Absolute magnitudes in {\it g}, (6), and {\it r}, (7), bands. Total luminosity in solar units in {\it g}, (8), and {\it r}, (9), bands.}
	\end{deluxetable*}

We derived the azimuthally averaged color profiles and, from images, the two-dimensional color maps. The mean $g-r$ values, reported in Table~\ref{tab:fit1d} and in Table~\ref{tab:fit1dIC}, are calculated up to the radii where the errors are less than $\sim 0.1-0.2$~mag.
Finally using stellar population synthesis models \citep{Ricciardelli2012,Vazdekis2012} with $\log Z/Z_{\odot}=0$ and a Kroupa initial mass function, and considering the average color which provides a stellar mass-to-light ratio ($\mathcal{M}/L$) in the {\it r} SDSS band \citep{Iodice2017a}, we estimate the total stellar mass (see Table~\ref{tab:fit1d} and Table~\ref{tab:fit1dIC}).

\subsection{The effect of PSF on the light distribution}
According to \citet[][and references therein]{deJong2008,Labarbera2012,Trujillo2016} it is important, especially for the brightest cluster galaxies (BCGs, hereafter), to account for the effect of the contamination from the galaxies bright cores and scattered light. This is done by mapping the Point Spread Function (PSF) out to a comparable radial distance of galaxy stellar halo.
We use the extended PSF derived for VST images by \citet[][Appendix~B]{Capaccioli2015}. The deconvolution was successfully applied at the bright galaxies in the VEGAS survery by \citet{Spavone2017b}  and it is based on the the Lucy-Richardson algorithm \citep{Lucy1974,Richardson1972}. A full description of the deconvolution method used for VST data will be presented in a forthcoming paper by Spavone et al. (in preparation).
The deconvolved azimuthally averaged surface brightness  {\it g} and {\it r}-band profiles for NGC~1533 are shown in Figure~\ref{fig:1533psf}. 
As expected \citep[see also][]{deJong2008}, in the seeing-dominated regions,  $R<1\arcsec$, the surface brightness profiles are brighter, of about $0.7-0.75$~mag~arcsec$^{-2}$ for both bands. In the outer regions, for $R\geq 1.5$ arcmin,  the profiles are steeper.

\section{Results}\label{sec:results}

\begin{figure*}
	\gridline{\fig{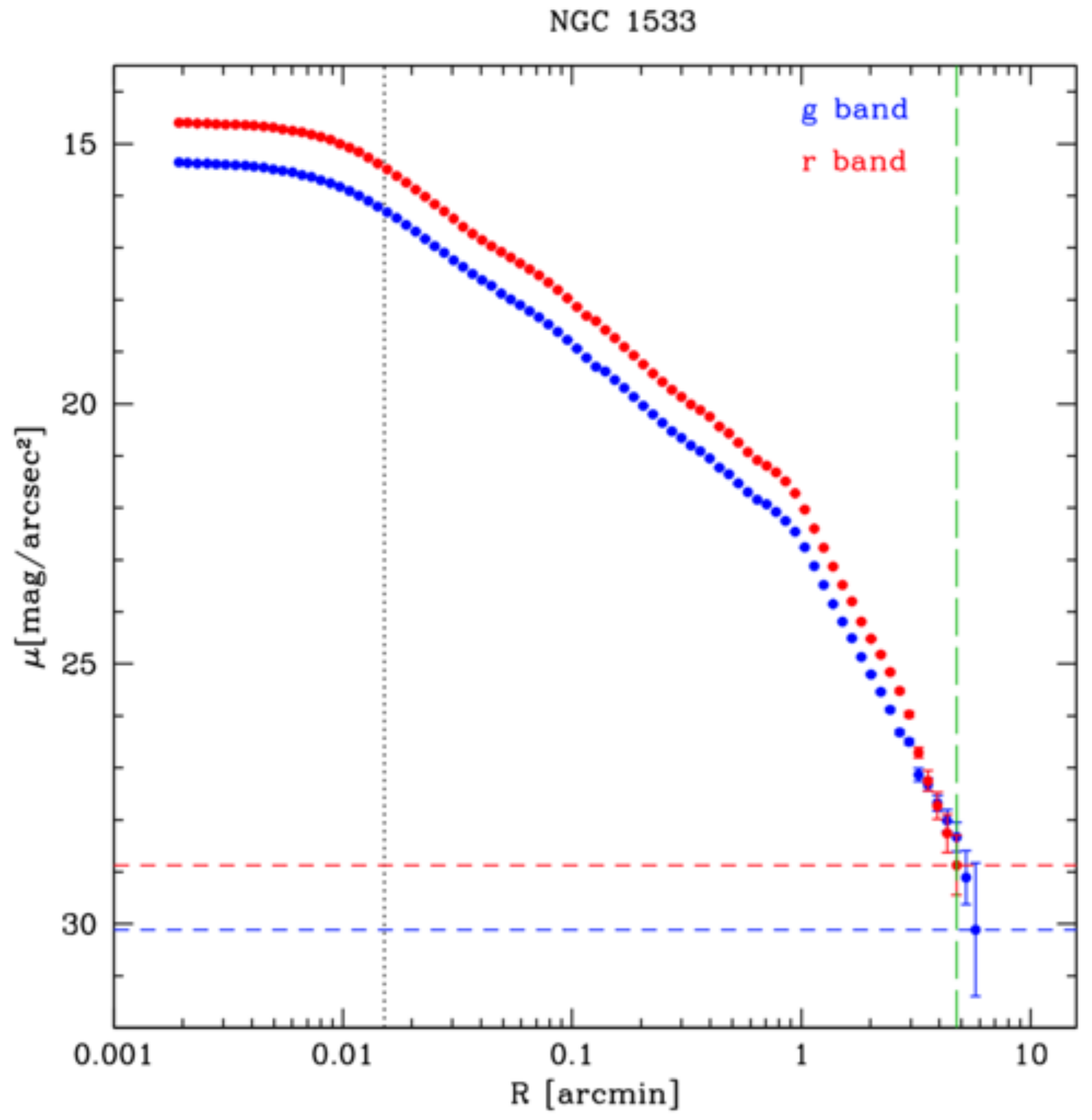}{0.449\textwidth}{}
		\fig{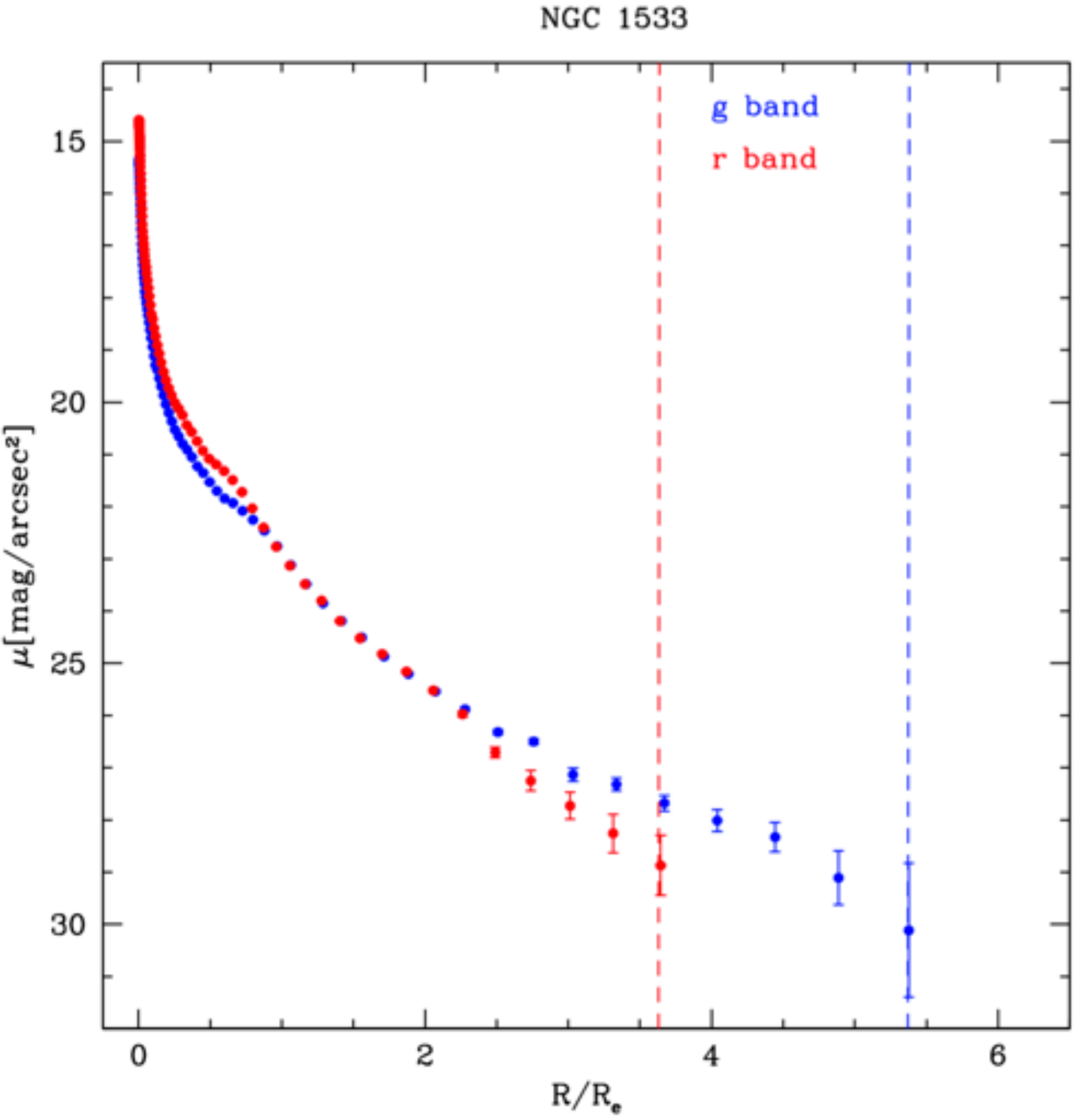}{0.45\textwidth}{}
	}
	\caption{{\it Left panel}: azimuthally averaged and PSF deconvolved surface brightness profile as a function of the logarithmic isophote semi-major axis; the black dotted line delimits the seeing dominated region. The red and blue dashed lines represent the faint surface brightness reached: $\mu_g=30.11$~mag~arcsec$^{-2}$ and $\mu_r =28.87$~mag~arcsec$^{-2}$; the green dashed line indicates relative radius $R=4\farcm75$. {\it Right panel}: averaged and PSF deconvolved surface brightness profile as a function of the isophote semi-major axis normalized to the effective radius. The red and blue dashed lines indicate the $R/R_e$ distances joined.}\label{fig:1533}
\end{figure*}

\begin{figure*}
	\plotone{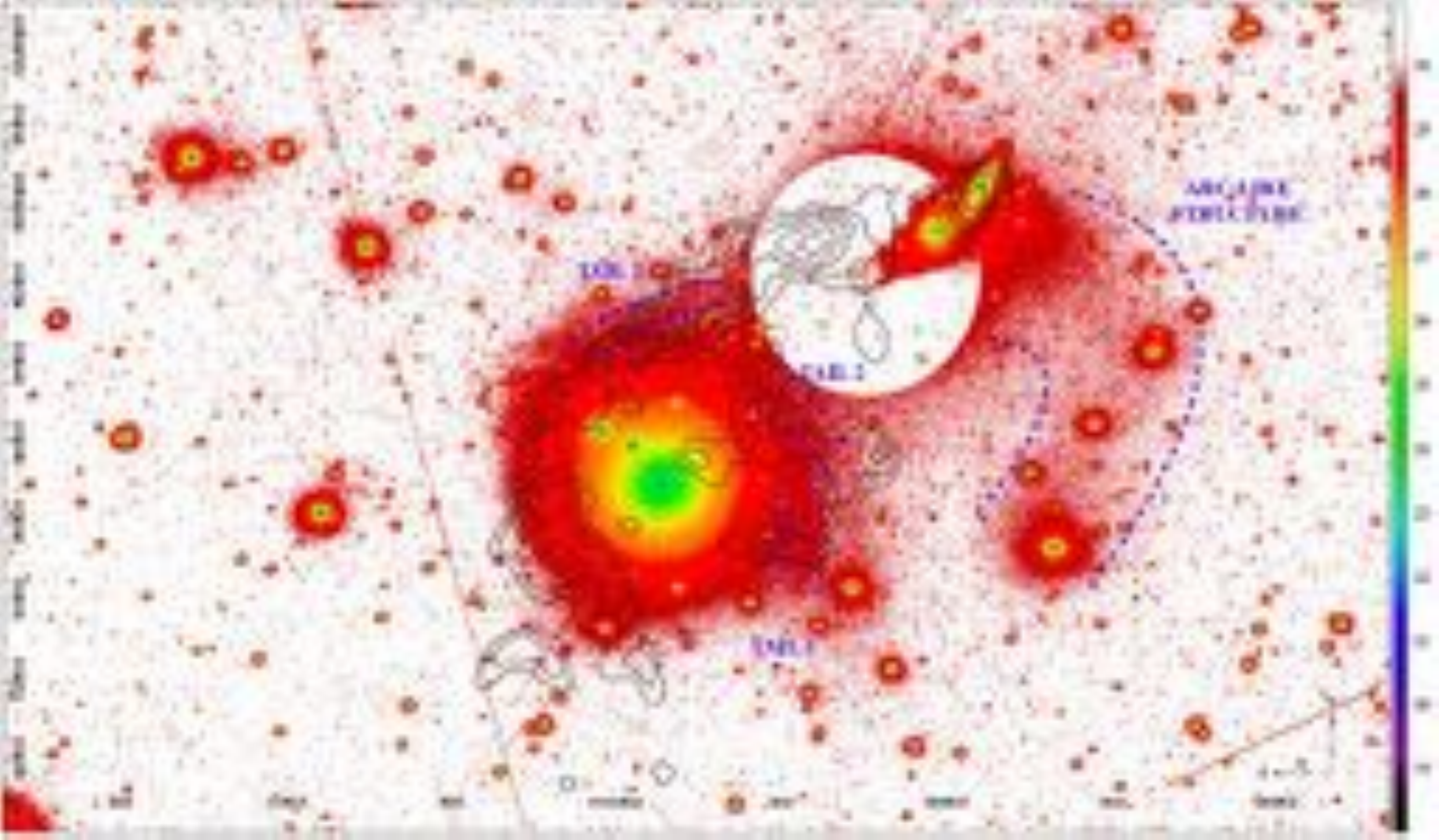}
	\caption{The image in surface brightness levels in {\it g} band of NGC~1533 system, with the HI map from the Australia Telescope Compact Array (ATCA) superimposed (black contours). The contours are $2.5$, $2.8$, $3.1$, $3.5$, $3.9$, $4.2\times10^{20}$~atoms~cm$^{-2}$ and have a resolution of $\sim1 \arcmin$ \citep{RyanWeber2003b}.  The spiral-like tails have a central distance of $229\farcs27$, $237\farcs02$ and $220\farcs34$ ($\sim21.19$, $21.90$, $20.36$~kpc), respectively for Tail1, Tail2 and Tail3; the arc-like structure is extended for $\sim5 \arcmin$ and it has a mean central distance of $\sim9 \arcmin$ ($\sim 49.90$~kpc). \label{fig:surbrig}}
\end{figure*}

OmegaCAM@VST deep photometry allows to reach the surface brightness limits of $\mu_g = 30.11 \pm 1.2$~mag~arcsec$^{-2}$ and $\mu_r = 28.87 \pm 0.6$~mag~arcsec$^{-2}$ at $R = 4\farcm75$ ($\sim 16.19$~kpc) for NGC~1533 (Figure~\ref{fig:1533}, left panel).
The VST image in the {\it g} band surface brightness levels is shown in Figure~\ref{fig:surbrig}.

In the following Subsections we describe the results obtained by mapping the light and color distributions out the unprecedented limits. In order to reduce the contamination of the light coming from the bright star CD$-56$~$854$, ($\alpha_{J2000}=4^h$~$9^m$~$15.03^s$ and $\delta_{J2000}=-56^{d}$~$01^{m}$~$48^{s}.45$) it has been modeled by using a Gaussian function and subtracted from both bands images; all the photometric analysis takes into account this, as it is shown in Figure~\ref{fig:vstcolcomposite} and in Figure~\ref{fig:surbrig}.

\subsection{The Faint Features in NGC~1533, IC~2038 and IC~2039 outskirts}\label{subsec:faint}

The deep images obtained for NGC~1533 with OmegaCAM@VST allow us to map the light distribution out to the galaxy outskirts and detect new faint features in these regions. 
In the {\it g} band, the outskirts of NGC~1533 appear asymmetric, i.e. light is elongated on the NE side. We found three faint $\mu_g \geq 28-30$~mag~arcsec$^{-2}$ spiral-like tails in the N-NW outskirt of the galaxy (see Figure~\ref{fig:surbrig}). In addition, we detect an extended ($\sim 5 \arcmin$) arc-like structure with $\mu_g \sim 28-30$~mag~arcsec$^{-2}$, protruding from the two interacting galaxies IC~2038 and IC~2039 on the West side.
The tails in the outskirts of NGC~1533 are still evident in {\it r} band image, while the faint arc-like structure on the West does not appear in this band.
This would be hint of a bluer color for this structure or, alternatively, it is comparable to the background level so it was accounted for in the polynomial fit of the sky fluctuations. 

Comparing the previous works on NGC~1533 \citep[e.g.][]{RyanWeber2003b,RyanWeber2004,Werk2008} and thanks to the deep VST image we are able to associate the optical counterpart to the HI gas.
Figure~\ref{fig:surbrig} shows the overlap of the deep VST {\it g} band image in surface brightness levels and the HI contours levels (where they are at column densities of $2.5$, $2.8$, $3.1$, $3.5$, $3.9$, $4.2\times10^{20}$~atoms~cm$^{-2}$). As it is anticipated by Bekki's simulations, a N-body/SPH numerical simulation, using the \citet{Fall1980} model, of a minor, unequal-mass merger \citep{RyanWeber2003b}, the surface brightness of stellar remnant is down to $\sim26-29$~mag~arcsec$^{-2}$. 
At these levels, $\mu_g \gtrsim28.5$~mag~arcsec$^{-2}$, we are able to trace the regions of the stellar envelope, which appears asymmetric and more elongated toward NE (in correspondence of Tail3). 
We find an overlap between the new faint features detected from the deep VST images and the HI distribution. It seems that most of the HI is within the stellar envelope of the galaxy on the region where it appears more elongated, so the light overdensity corresponds to the HI overdensity. The HI on West side is also concentrated on the two spiral-like tails (Tail1 and Tail2) in the stellar envelope on the galaxy. 

\subsection{Light and color distribution}\label{}

\begin{figure}
	\plotone{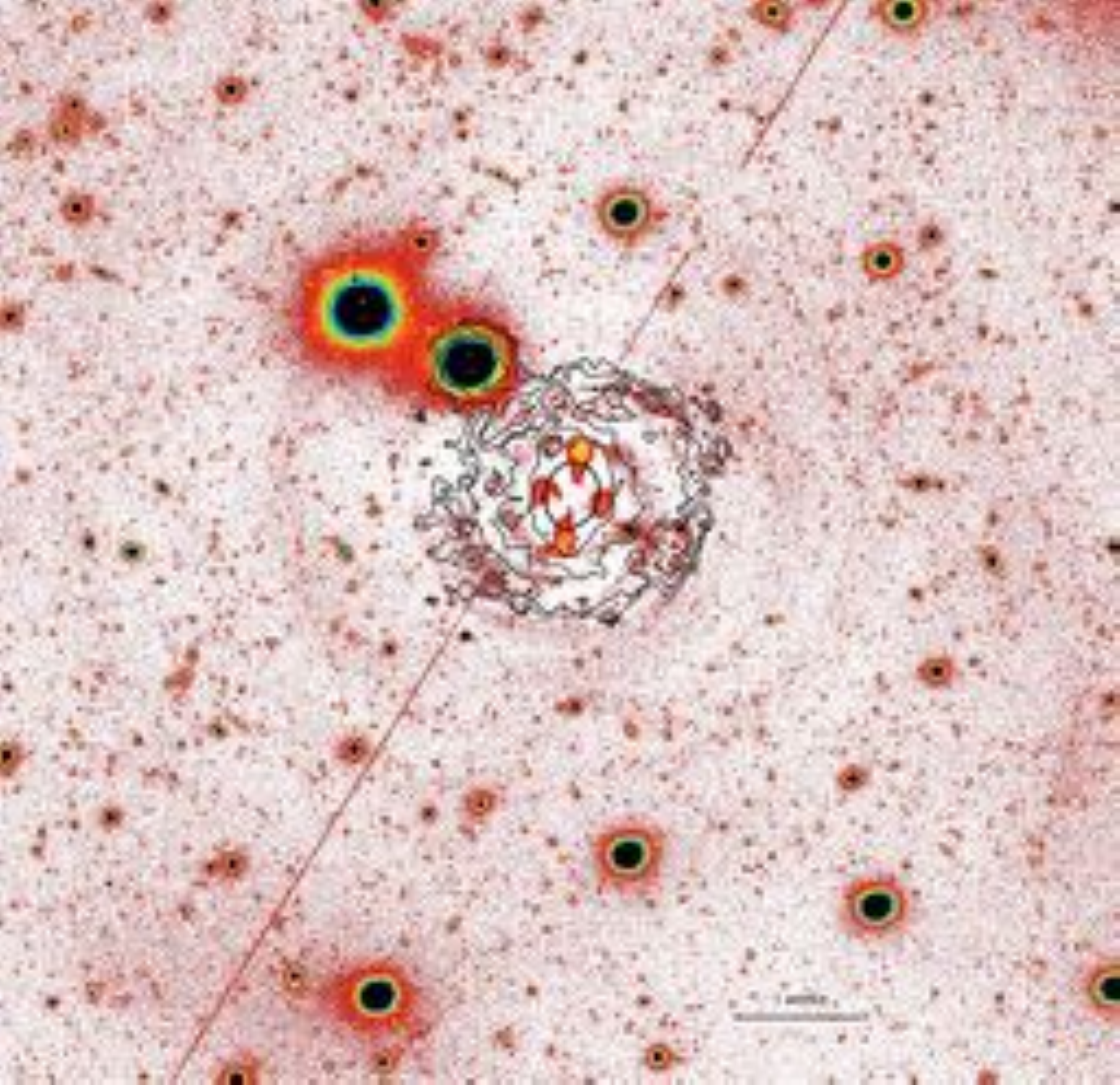}
	\caption{Residual image ($8\farcm30 \times 7\farcm95$) from VST {\it g} band of NGC~1533 inner region with the contours (levels are $25.26$ (bold), $25.85$, $26.10$, $26.29$, $26.41$ (dashed)~mag~arcsec$^{-2}$) obtained from the {\it W2}  filter image {\it Swift}-{\tt UVOT} \citep{Rampazzo2017}. \label{fig:bmodelW2}}
\end{figure}

\begin{figure}
	\plotone{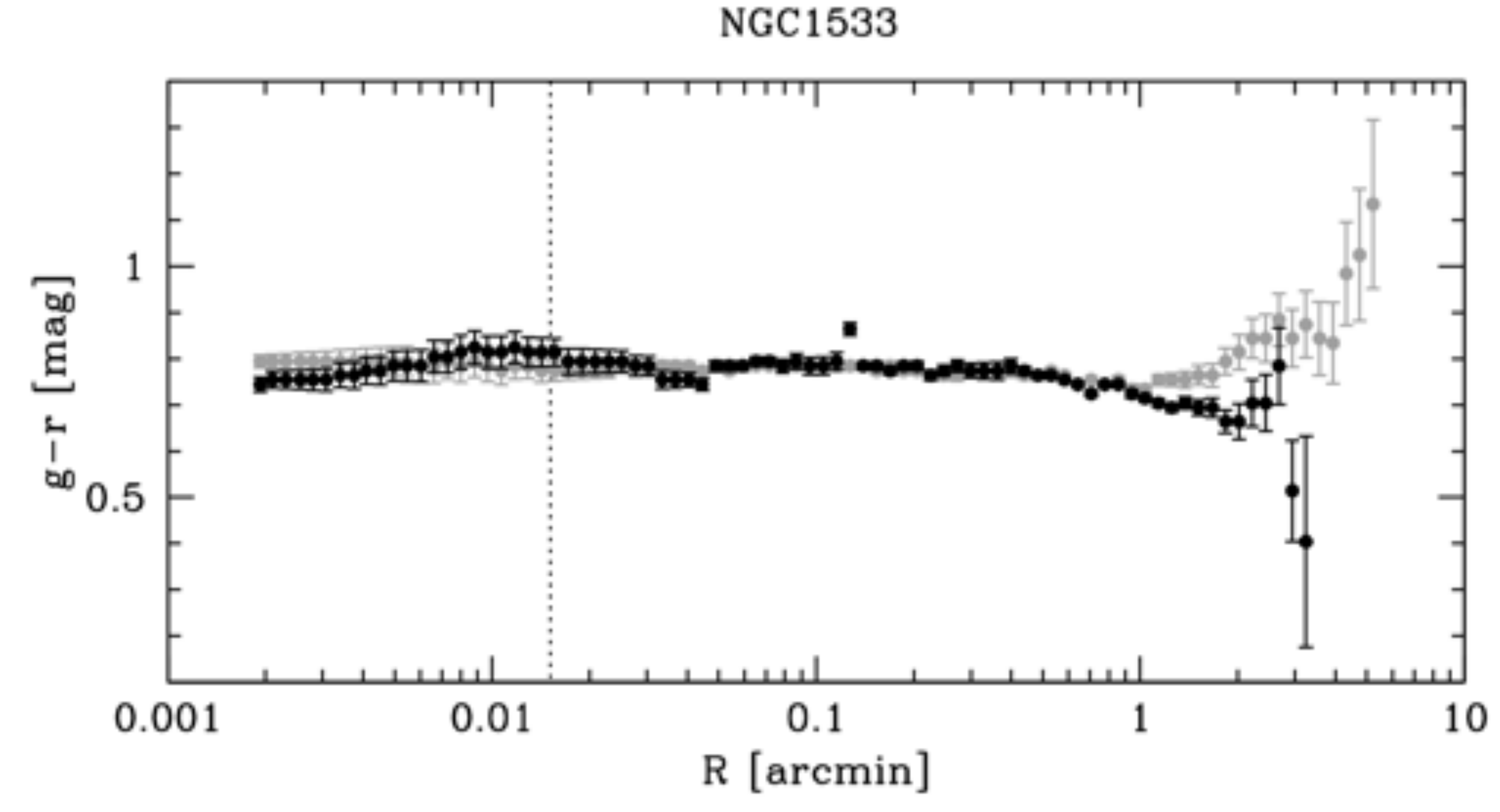}
	\plotone{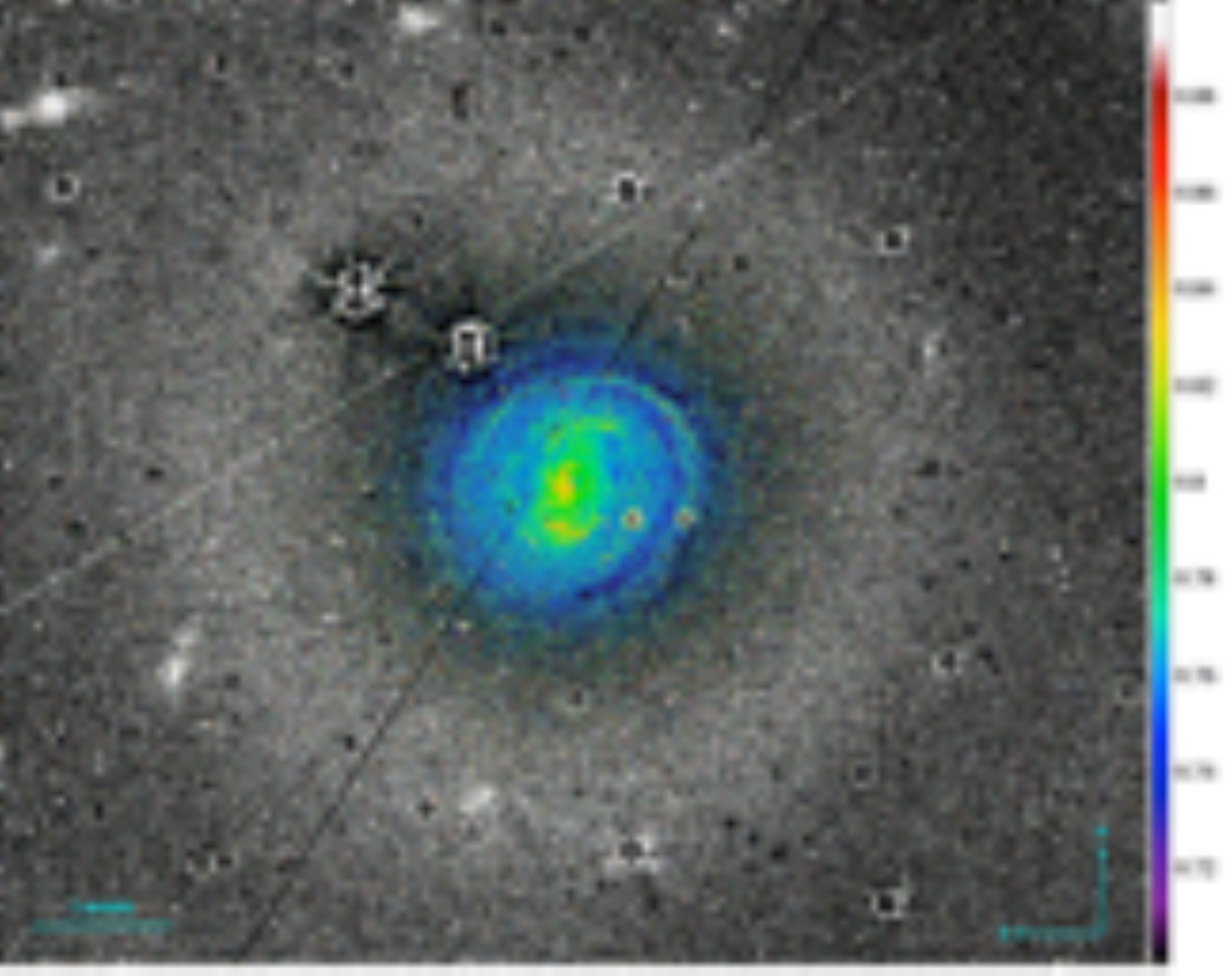}
	\caption{{\it Top panel}: azimuthally averaged extinction corrected color profile as a function of the logarithmic isophote semi-major axis; original profile (gray dots) and deconvolved profile (black dots). The black dotted line delimits the seeing dominated region. {\it Bottom panel}: 2D color map centered in NGC~1533 made using the deconvolved images.}\label{fig:1533col}
\end{figure}

{\it NGC~1533 $-$} 
From the isophote fit we made the 2-dimensional model of the galaxy in the {\it g} band,  by using the IRAF task \textsc{bmodel}, and we derived the residual image, shown in Figure~\ref{fig:bmodelW2}. Residuals clearly revealed an optical previously undetected arm-like structure, which is brighter in the inner regions, $30\arcsec \leq R \leq 
60\arcsec$, but still evident at larger radii, where the faint tails on West side (see Figure~\ref{fig:surbrig}) also stand out. 
The same structure is also detected in the 2D $g-r$ color map  (see Figure~\ref{fig:1533col}, bottom panel). The spiral-like pattern resemble a spiral disk where redder arms ($g-r \sim 0.8$~mag) are interloped by bluer arms ($g-r \sim 0.72-0.78$~mag) made of dust and young stars. 

The azimuthally averaged surface brightness profiles in the {\it g} and {\it r} bands are shown in Figure~\ref{fig:1533} (left panel). From a visual inspection, the galaxy is made by the inner luminous bulge, followed by a lens and bar plateau, and the outer ring-lens, disk and stellar envelope component. The multi-component fit, to set the scale of the most luminous components dominating the light distribution, is performed and described in Section~\ref{sec:light}. 

The mean color value, estimated from the azimuthally averaged extinction corrected \citep{Schlafly2011} and PSF deconvolved $g-r$ color profile (Figure~\ref{fig:1533col}, top panel), up to $\sim 5 \arcmin$, is $0.77 \pm 0.05$~mag (see also Table~\ref{tab:fit1d}). 

The effect of the PSF wings at large radial distances can alters the color profile of galaxies, showing the so called {\it red halo phenomenon} \citep[][and references therein]{Michard2002,Zackrisson2006,Labarbera2012}.
It depends more on the observational bands: redder is the band greater is the red gradient in the galaxy outskirts color profile. Figure~\ref{fig:1533col} (top panel) shows the color profile of NGC~1533 before and after the PSF deconvolution. It is clear that in the outer regions, $R>1\arcmin$, the mean color changes from $\sim0.95$~mag to $\sim0.46$~mag.

Finally, for face-on and round galaxies, as NGC~1533, the PSF deconvolution doesn't modify the geometrical parameters of the outer isophotes because the PSF effects is larger for bright, smaller and edge-on galaxies \citep{Michard2002,deJong2008}.

Using the growth curve, we derived the total magnitudes and effective radii ($R_e$), and hence the absolute magnitudes and the total luminosity in solar units are estimated, in the {\it g} and {\it r} bands respectively (see Table~\ref{tab:results1}). The effective radii in {\it g} and {\it r} bands are $R_e \sim 5.93$~kpc and $R_e \sim 7.23$~kpc, respectively. Therefore, the surface brightness profiles extend out to $\sim 5 R_e$ and $\sim 4 R_e$, respectively (see Figure~\ref{fig:1533}, right panel).

\begin{figure}
	\plotone{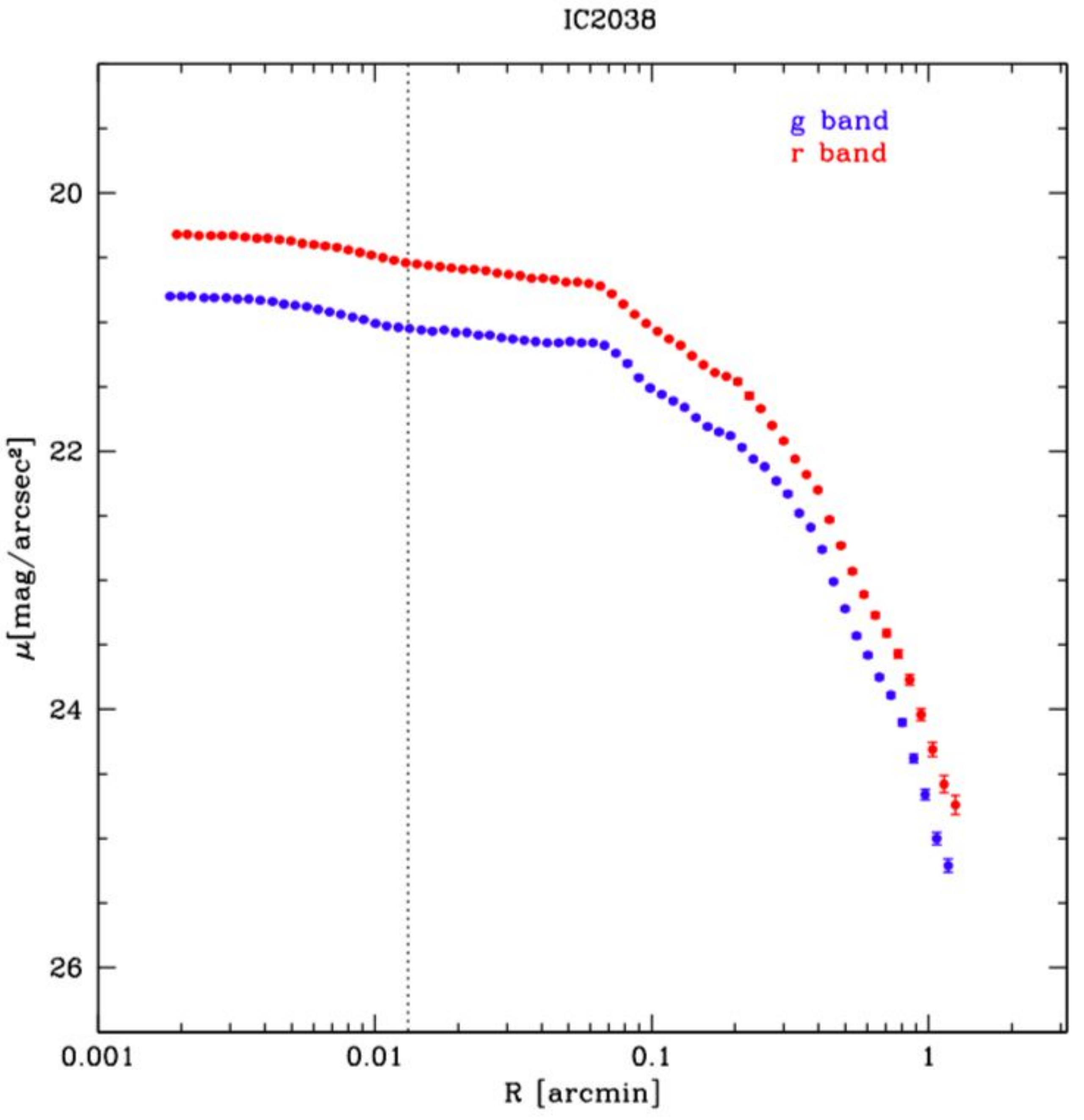}
	\plotone{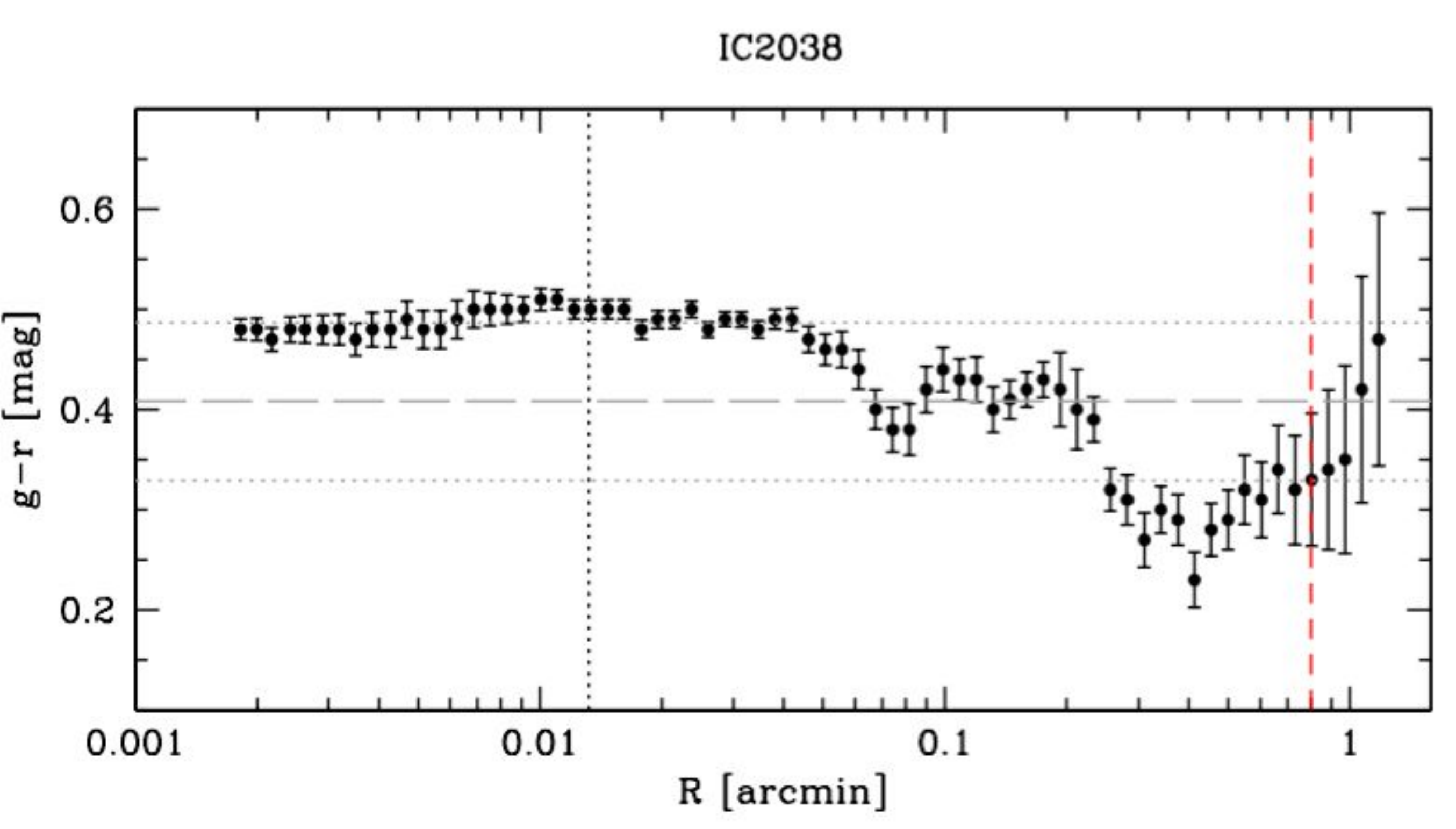}
	\caption{{\it Top panel}: azimuthally averaged surface brightness profile as a function of the logarithmic isophote semi-major axis; {\it g} band image, blue dots, and {\it r} band image, red dots. {\it Bottom panel}: azimuthally averaged color profile as a function of the logarithmic isophote semi-major axis. The red dashed vertical line represents the bound of the region where the mean color value, horizontal dashed gray line, has been estimated.\label{fig:2038}} 
\end{figure}

\begin{figure}
	\plotone{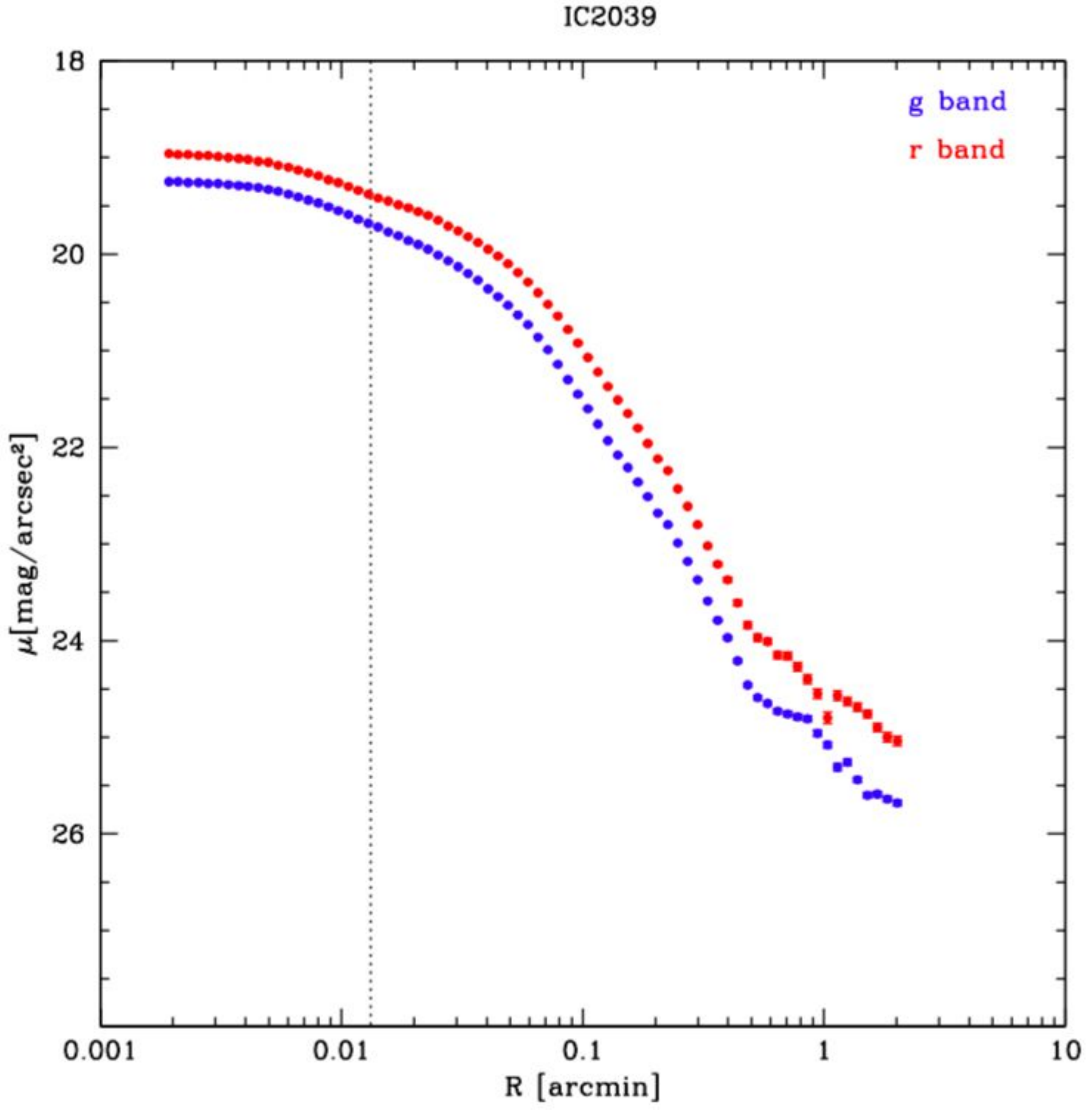}
	\plotone{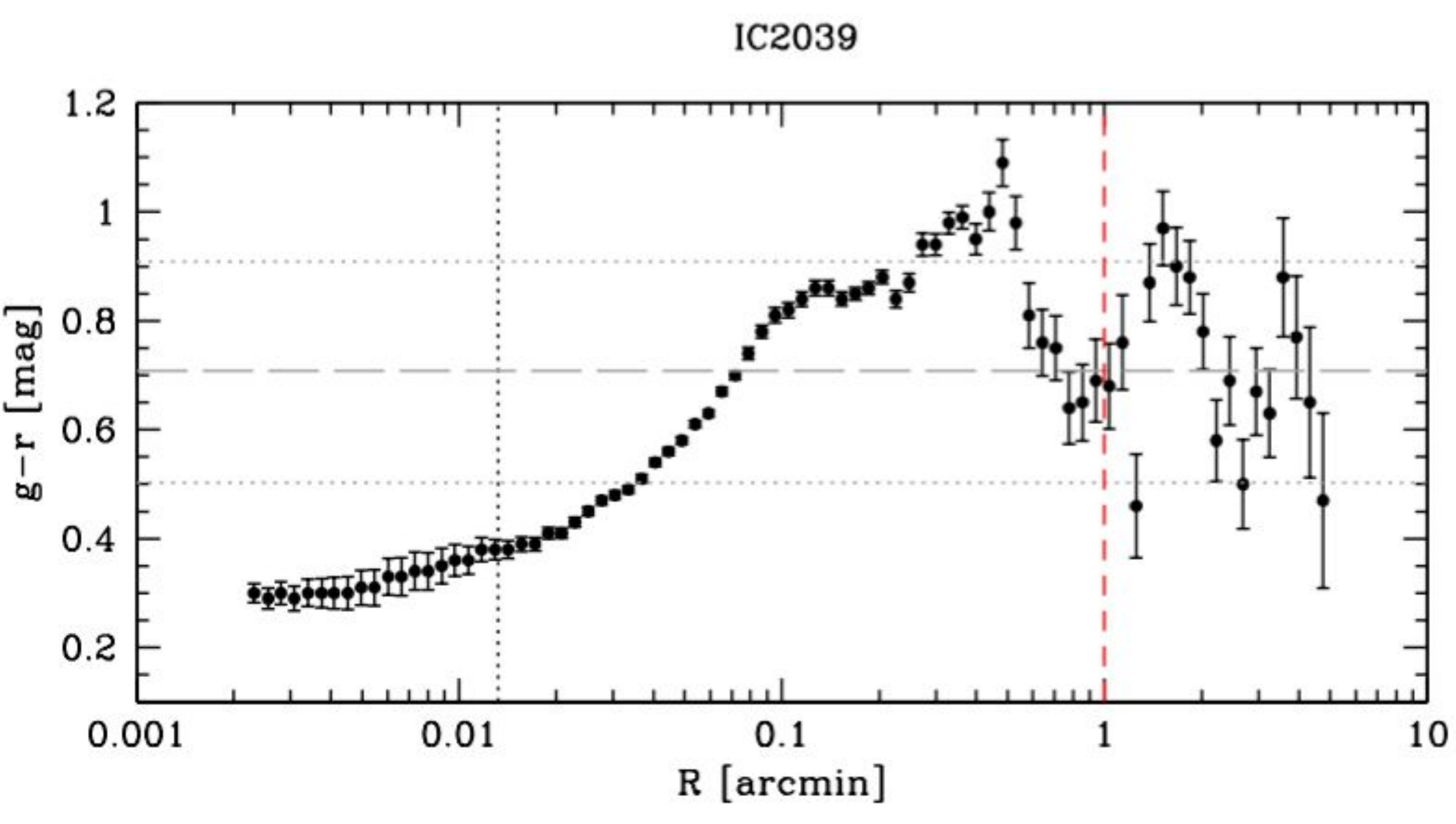}
	\caption{The same as Figure~\ref{fig:2038} for IC~2039}.\label{fig:2039}
\end{figure}

{\it IC~2038 and IC~2039 $-$} 
The azimuthally averaged surface brightness profiles derived for IC~2038 and IC~2039 are shown in Figure~\ref{fig:2038} (top panel) and Figure~\ref{fig:2039} (top panel), respectively.
Figure~\ref{fig:2038} (bottom panel) shows the azimuthally averaged color profile of IC~2038, Figure~\ref{fig:2039} (bottom panel) of IC~2039 (see also Table~\ref{tab:fit1dIC}).
The mean color value of IC~2038 is $g-r=0.41\pm0.08$~mag, which is consistent with the typical colors observed for late-type galaxies \citep{Strateva2001}. 
It shows a consistent decrease, of about $0.04$ mag out to $R\sim4\farcs1$, in the region where spiral arms dominate the galaxy structure.
IC~2039 is redder, with $g-r=0.71\pm0.20$~mag (see also Table~\ref{tab:fit1dIC}), which is comparable with the typical colors observed for early-type galaxies \citep{Labarbera2012}.  
The color profile shows a blue dip in the inner regions, where $0.38 \leq g-r \leq 0.61$~mag for $0\farcs9\leq R \leq 3\farcs2$. 
At larger radii, the color profile has a scattered trend in a wide range of values, $g-r \sim 0.6 - 1.1$ mag out to $\simeq 26\arcsec$.

\subsection{Multi-component fit of the light distribution}\label{sec:light}
In order to define the scales of the main components that dominate the light distribution in NGC~1533, we performed a multi-component fit to reproduce the surface brightness profile, as done for the BCGs at the center of the groups and clusters 
\citep{Seigar2007,Donzelli2011,Arnaboldi2012,Cooper2013,Huang2013,Cooper2015,Iodice2016,Rodriguez2016,Spavone2017b,Spavone2018,Iodice2018}. 
BCGs consist of at least two components, the bright spheroidal body and the outer and very extended,  moderately flattened, stellar envelope.  

Even if the isophotal analysis has shown that the inner structure of NGC~1533 is quite complex, including a bar and a lens, 
we adopted a simple model that should reproduce the most luminous components of the galaxy.
Therefore, the residuals would point out the main subcomponents hidden in the light distribution. 

As starting point, we performed a least-squares fits using a Levenberg-Marquardt algorithm \citep{Seigar2007,Spavone2017b} of the azimuthally averaged surface brightness profile  {\it g} band. Our model is made up by a composition of  S\'{e}rsic power law \citep{Sersic1968}, $\mu (R) = \mu_e + k(n) \left[ \left( \frac{R}{r_e} \right) ^{1/n} -1 \right]$, where {\it R} is the galactocentric radius, $r_e$ the effective radius, $\mu_e$ the surface brightness at the effective radius, $k(n) = 2.17n-0.355$ \citep{Caon1993}, and exponential law  \citep{Freeman1970}, $\mu (R) = \mu_0 + 1.086 \times \frac{R}{r_h}$, where {\it R} is the galactocentric radius, $r_h$ the exponential scale length and $\mu_0$ the central surface brightness (see Table~\ref{tab:fit1d}, Table~\ref{tab:fit1dIC} and Figure~\ref{fig:fit1d}).
The total magnitude obtained from fit is $\sim 0.5$~mag brighter than that measured from the growth curve. Such a difference is due to the exclusion of the central seeing-dominated regions (R$\leq 0\farcs79$) from the fit.

Considering the state-of-art of studies on the multi-component fit of the light distribution \citep[see e.g.][and references therein]{Spavone2017b,Spavone2018} it is possible to derive the relative contribution of the accreted component with respect to the total galaxy light, $f_{h,T}$, and the total accreted mass fraction (see Table~\ref{tab:fit1d} and Table~\ref{tab:fit1dIC}).

The structural parameters obtained by the one-dimensional fit (see Table~\ref{tab:fit1d} and Table~\ref{tab:fit1dIC}) are used as starting guesses for the two-dimensional fit, performed by using GALFIT\footnote{GALFIT is a data analysis algorithm that fits 2D analytic functions of galaxies and point sources directly to digital images. A detailed description of the GALFIT technique can be found in \citet{Peng2002}, with new features found in \citet{Peng2010} https://users.obs.carnegiescience.edu/peng/work/galfit/galfit.html}.  
In order to improve the quality of the resulting model and residual image we have added in the input file the mask of all the bright sources and background/foreground objects, and the mean value of the residual background fluctuations.

For NGC~1533 the best fit is made by one S\'ersic law and two exponential laws, see Table~\ref{tab:fit}. 
The first function accounts for the bulge, the inner component; the second one approximate the lens plus inner disk, and the third models the outer asymmetric envelope (see Figure~\ref{fig:fit2d}, top middle panel). The stellar envelope starts to dominate the light for $R \geq 2 \arcmin$ ($\sim 6.82$ kpc) and has an {\it r} band surface brightness in the range $25 - 29$~mag~arcsec$^{-2}$ (see Figure~\ref{fig:fit1d}).
The 2-dimensional fit is consistent with that obtained on the azimuthally-averaged profile (see Table~\ref{tab:fit}, Table~\ref{tab:fit1d} and Table~\ref{tab:fit1dIC}) and results (model and residuals) are shown in Figure~\ref{fig:fit2d}. 
The 2-dimensional model (see middle panel of Figure~\ref{fig:fit2d}) confirms that the outskirts of NGC~1533 are asymmetric, showing that isophotes are more elongated in the NE direction.
The residual image reports the bar and the spiral-like structure, which has just been revealed by the two-dimensional color map (see Figure~\ref{fig:fit2d}, top right panel).

Despite of what happens for NGC~1533, in order to have a better estimation of the light distribution between the two interacting galaxies, IC~2038 and IC~2039, and to have a residual image consistent with the system, we contemporary modeled both galaxies in order to take into account the contribution of light of each of them in the overlapping regions (see Figure~\ref{fig:fit2d}, bottom middle panel). 
The light profiles of IC~2038 and IC~2039 are best fitted by a single S\'{e}rsic law. 
The best fit parameters are reported in Table~\ref{tab:fit} and the best 2-dimensional models are shown in Figure~\ref{fig:fit2d} (lower-middle panel).
In the residual image appears an elongated structure in the direction of NGC~1533$-$IC~2038 and the light overdensity in the West side of IC~2038 (see Figure~\ref{fig:fit2d}, lower-right panel, and Figure~\ref{fig:surbrig}).

\section{Discussion: tracing the build-up history of the NGC~1533 cloud}\label{sec:discussion}

In this work we have analyzed the deep VST images, in the {\it g} and {\it r} bands, of the Dorado group, centered on the brightest group member NGC~1533. 
We studied the morphology, light and color distribution of the galaxy members in the VST fields.
The large integration time and the wide field of view of OmegaCAM@VST allow us to map the surface brightness down to 
$\mu_g \simeq 30.1$~mag~arcsec$^{-2}$ and $\mu_r \simeq 28.9$~mag~arcsec$^{-2}$ for NGC~1533 ad out  $4 R_e$, therefore out to regions of the stellar halo.
The main and new results based on the VST deep imaging are:
\begin{itemize}
	\item the optical counterpart of the inner spiral-like features detected in UV by \citet{Marino2011b} with {\tt GALEX} and by \citet{Rampazzo2017} with {\it Swift} (see Figure~\ref{fig:bmodelW2});
	\item the detection of  three faint ($\mu_g \geq 28-30$~mag~arcsec$^{-2}$) spiral-like tails in the N-NW outskirt of the galaxy (see Figure~\ref{fig:surbrig}) and an extended ($\sim 5 \arcmin$) arc-like structure with $\mu_g \sim 28-30$~mag~arcsec$^{-2}$, protruding from the two interacting galaxies IC~2038 and IC~2039 on the West side;
	\item by fitting the extended light profile in the {\it r} band, we found that the stellar envelope of NGC~1533 has an exponential profile, it extends out to about $4.75$~$\arcmin$ ($\sim 16$~kpc) and spans the range $\mu_r = 25 - 29$~mag~arcsec$^{-2}$ (see Section~\ref{sec:light}).
\end{itemize}

In the following sections we discuss the implications from the analysis performed in this work  on the  building-up of the Dorado group of galaxies.
In detail, we address the main conclusions on the structure of NGC~1533, from the inner regions to the faint outskirts and on the ongoing interactions between the group members, in turn.  

\subsection{The inner structure of NGC~1533: the relics of a past merger}

The 2D $g-r$ color map of NGC~1533 (Figure~\ref{fig:1533}, bottom right panel) revealed an optical previously undetected arm-like structure in the inner regions, $30\arcsec \leq R \leq 60\arcsec$. The same structure is also evident in the residual image obtained by subtracting the two-dimensional model, obtained from the isophote fit from the original image of NGC~1533 in the {\it g} band (see Figure~\ref{fig:bmodelW2}). 
The spiral pattern with redder colors ($g-r \sim0.8$~mag) on an average bluer ($g-r \sim0.72-0.76$~mag) regions
would resemble a dusty disk. In the same regions, \citet{Rampazzo2017} detected ultra-violet (UV) emissions, which appear as ring or lens \citep[see also][]{Comeron2014}.
Figure~\ref{fig:bmodelW2} shows the residual image of NGC~1533 and the contours levels obtained from the {\it W2} filter image from {\it Swift}-{\tt UVOT} \citep{Rampazzo2017}, where it clearly appears that the ultraviolet arc-like and spots emission regions correspond to the optical arm-like structures. 
The UV emission detected in the central regions of the galaxy was suggested to be a sign of a recent star formation \citep{Rampazzo2017}. 
Therefore, this coincidence between the optical and color features detected from VST images and the UV emissions 
would suggest a merger event that shaped the spiral structure and triggered star formation in the  central regions  of NGC~1533. 
\begin{longrotatetable}
	\begin{splitdeluxetable*}{lccccccccccccBcccc}
		\tabletypesize{\scriptsize}
		\tablecaption{Best fitting structural parameters for a three-component fit of NGC~1533}\label{tab:fit1d}
		\tablecolumns{17}
		\tablenum{4}
		\tablewidth{0pt}
		\tablehead{
			\colhead{Filter}
			&\colhead{r$_{e1}$} 
			&\colhead{n$_1$} 
			&\colhead{$\mu_{e1}$} 
			&\colhead{m$_{\text{tot,1}}$} 
			&\colhead{r$_{e2}$} 
			&\colhead{n$_2$}
			&\colhead{$\mu_{e2}$}
			&\colhead{m$_{\text{tot,2}}$} 
			&\colhead{r$_{0}$} 
			&\colhead{$\mu_0$}
			&\colhead{m$_{\text{tot,3}}$}
			&\colhead{$f_{h,T}$}
			& \colhead{$g-r$}
			& \colhead{$(M/L)_{r}$} 
			& \colhead{$\mathcal{M}^{*}_{tot}$}
			& \colhead{$\mathcal{M}^{*}$$_{tot}$~$_{acc}$}\\
			\colhead{} 
			&\colhead{[arcsec]} 
			&\colhead{}
			&\colhead{[mag~arcsec$^{-2}$]} 
			&\colhead{[mag]}
			&\colhead{[arcsec]} 
			&\colhead{}
			&\colhead{[mag~arcsec$^{-2}$]} 
			&\colhead{[mag]}
			&\colhead{[arcsec]} 
			&\colhead{[mag~arcsec$^{-2}$]}
			&\colhead{[mag]}
			&\colhead{}
			&\colhead{[mag]} 
			&\colhead{$\left[ \mathcal{M}_{\odot}/L_{\odot} \right]$}
			&\colhead{$\left[ \mathcal{M}_{\odot} \right]$}
			&\colhead{$\left[ \mathcal{M}_{\odot} \right]$}\\
			\colhead{(1)} 
			&\colhead{(2)}
			&\colhead{(3)}
			&\colhead{(4)}
			&\colhead{(5)}
			&\colhead{(2)}
			&\colhead{(3)}
			&\colhead{(4)}
			&\colhead{(5)}
			&\colhead{(2)}
			&\colhead{(4)}
			&\colhead{(5)}
			&\colhead{(6)}
			&\colhead{(7)}
			& \colhead{(8)}
			& \colhead{(9)}
			& \colhead{(10)}
		}
		\startdata
		{\it g} & $5.04\pm0.06$ & $2.34\pm0.68$ &  $18.88\pm0.53$ & $11.98$ & $39.48\pm8. 97$ & $1.20\pm0.16$ & $21.97\pm0. 88$ & $10.60$ & $109.86\pm1.5$ &$25.58\pm0. 01$&$13.38$&$79\%$&$0.77\pm0.05$&$2.88$&$1.27 \times 10^{11}$&$0.99 \times 10^{11}$\\
		{\it r}  & $5.00\pm0.40$ & $2.29\pm0.10$ & $18.05\pm0.21$ & $11.17$ & $39.02\pm4. 22$ & $1.20\pm0.10$ & $21.19\pm0. 01$  & $9.85$ & $57.92\pm0.24$ & $23.71\pm0. 64$ & $12.9$ & $78\%$&&&&\\
		\enddata
		\tablecomments{(1) Filters. Best fit parameters of the PSF deconvolved light profiles decomposition: (2) effective and scald radius; (3) S\'{e}rsic index, (4) effective and central surface brightness of the different components and (5) total magnitude of the individual component. (6) Total accreted mass fraction derived from our fit. (7) Averaged extinction-corrected and PSF deconvolved $g-r$ color value and relative mass-to-light ratio in {\it r} band (8). Total stellar mass (9) and total accreted stellar mass fraction derived from the three-component fits (10) in {\it r} band.}
	\end{splitdeluxetable*}
\end{longrotatetable}

Recently \citet{Mazzei2014a} explain the inner structure of NGC~1533 within a major merger; NGC~1533 could arise from a merger between two halos with mass ratio 2:1 and perpendicular spins. The simulations reproduce the inner ring/arm-like structure visible in the UV bands and the entire SED of the galaxy. In this case the gas belongs to the merger remnant halo which could feed the SF. \citet{Thom2012} found that there exist conspicuous gas reservoirs around ETGs, similarly to late-types. 

\subsection{Stellar halo component of NGC~1533}\label{subsec:halo}

In this section we attempt to trace the building up history of the stellar halo of NGC~1533 by comparing the observed properties with some theoretical predictions. 
The stellar halos in galaxies are made of stars accreted during the merging events in the form of  a ``relaxed'' component,  which is superimposed to the stars formed in-situ, and the outer and diffuse stellar envelope that is dominated by the ``unrelaxed'' accreted material \citep{Cooper2013,Cooper2015}. Cosmological simulations of galactic stellar halo formation by the tidal disruption of accreted material are able to give a complete set of ``observables'' (as structure, density profiles and metallicity) that can be compared with real data. In particular, \citet{Cooper2010} made a detailed analysis of the properties for six simulated stellar halos. They found that the assembly history  is made either by a gradual accretion of several progenitors with roughly equal mass  or by the accretion of one or two systems. The resulting morphology of the stellar halo is different in the two cases \citep[see Fig.6 in][]{Cooper2010}.  The stellar halos built up by the gradual accretion are the most extended, out to $70-100$~kpc, and have several streams, shells  and other irregular structures, which are more evident at larger radii. On the contrary, the accretion of one or two massive satellites generates smaller stellar halos with a strong central concentration. All features span a surface brightness range $24-35$~mag~arcsec$^{-2}$ in the {\it V} band ($\sim 23-35$~mag~arcsec$^{-2}$ in the {\it r} band). 
The shape of the density profile is also different in the two cases, the halos formed by many progenitors are steeper and well fitted by a S\'{e}rsic profile with $n\sim1$, than those formed by  few progenitors

The deep VST images for  NGC~1533 allow us to map the faint regions of the stellar envelope  out to $10 \arcmin$ ($\sim 34.09$~kpc). 
This component starts to dominate the light for $R\geq2 \arcmin$ ($\sim 6.82$~kpc) and has an {\it r} band surface brightness in the range  $25-29$~mag~arcsec$^{-2}$ (see Section~\ref{subsec:faint}). 
Therefore, at these large distances and surface brightness levels we are able to make a direct comparison between the observed properties and those from theoretical predictions described above. 

The stellar envelope of NGC~1533 appears quite diffuse and asymmetric, being more elongated towards NE (see Figure~\ref{fig:surbrig}).
We have detected three faint tails, one on the North and two others on the West side (see Figure~\ref{fig:surbrig} and Section~\ref{subsec:faint}), which have a surface brightness of $\sim 28-30$~mag~arcsec$^{-2}$.
By fitting the azimuthally averaged surface brightness profile of NGC~1533, the outer component, mapping the stellar envelope, is well reproduced by an exponential function (see Section~\ref{sec:light} and Figure~\ref{fig:fit1d}).
The stellar envelope  has a spheroidal form and this is highlighted by the variations in position angle ($90^{\circ} \leq$P.A.$\leq 275^{\circ}$), ellipticity ($0.025 \leq \epsilon \leq 0.1$) and color profile (mean color $g-r \simeq 0.46$~mag) in the outer regions.
It is rounder than the inner components and twisted (see Figure~\ref{fig:geompar} and Appendix~\ref{apx:A}), and has bluer colors (see Figure~\ref{fig:1533col} and Section~\ref{sec:results}).

Compared to the simulated halos presented by \citet{Cooper2010}, the stellar halo in NGC~1533 has a similar morphology, surface brightness levels and light distribution to that of  the Aq-A model (shown in Figure~$6$ of that paper). The Aq-A halo results from the  gradual accretion of many progenitors. The stellar mass of the simulated Aq-A galaxy is about $2 \times 10^{10}$~$\mathcal{M}_{\odot}$, which is one order smaller than the stellar mass in NGC~1533 (see Section~\ref{sec:results}).

It appears with several shells and arc-like tails in the surface brightness range $27-34$~mag~arcsec$^{-2}$ in the {\it V} band, which corresponds to an r-band magnitude of $26-30$~mag~arcsec$^{-2}$, fully consistent with the range of surface brightness in the halo of NGC~1533, and it is characterized by an exponential light profile, as also found in NGC~1533.

A further support to the hypothesis  that the outskirts of NGC~1533 formed by the gradual accretion of minor mergers could come by comparing the average colors of these regions with those observed for the dwarf galaxies members of the group. Therefore, from \citet{Ferguson1990} we selected the dwarf galaxies in the Dorado group and we derived the mean $g-r$ colors (see Table~\ref{tab:dwarfs}). For all of them, the $g-r \sim 0.88\pm0.33$, which is consistent with the average $g-r$ color for the envelope in NGC~1533, i.e. at $R\geq1.5$~$\arcmin$ (see Figure~\ref{fig:1533col}). So dwarf galaxies of comparable size and luminosity could be considered as small satellites that should be accreted by NGC~1533 stellar envelope.

\begin{splitdeluxetable*}{lcccccccccBlcccc}
	\tablecaption{Best fitting structural parameters for a two-component fit\label{tab:fit1dIC}}
	\tablecolumns{18}
	\tablenum{5}
	\tablewidth{0pt}
	\tablehead{
		\colhead{Galaxy}
		&\colhead{Filter}
		&\colhead{r$_{e1}$} 
		&\colhead{n$_1$} 
		&\colhead{$\mu_{e1}$} 
		&\colhead{m$_{\text{tot,1}}$} 
		&\colhead{r$_{0}$} 
		&\colhead{$\mu_0$}
		&\colhead{m$_{\text{tot,3}}$} 
		&\colhead{$f_{h,T}$}
		&\colhead{Galaxy}
		& \colhead{$g-r$}
		& \colhead{$(M/L)_{r}$} 
		& \colhead{$\mathcal{M}^{*}_{tot}$}
		& \colhead{$\mathcal{M}^{*}$$_{tot}$~$_{acc}$}\\
		\colhead{} 
		&\colhead{} 
		&\colhead{[arcsec]} 
		&\colhead{}
		&\colhead{[mag~arcsec$^{-2}$]} 
		&\colhead{[mag]}
		&\colhead{[arcsec]} 
		&\colhead{[mag~arcsec$^{-2}$]}
		&\colhead{[mag]}
		&\colhead{}
		&\colhead{} 
		&\colhead{[mag]} 
		&\colhead{$\left[ \mathcal{M}_{\odot}/L_{\odot} \right]$}
		&\colhead{$\left[ \mathcal{M}_{\odot} \right]$}
		&\colhead{$\left[ \mathcal{M}_{\odot} \right]$}\\
		\colhead{} 
		&\colhead{(1)} 
		&\colhead{(2)}
		&\colhead{(3)}
		&\colhead{(4)}
		&\colhead{(5)}
		&\colhead{(2)}
		&\colhead{(4)}
		&\colhead{(5)}
		&\colhead{(6)}
		&\colhead{} 
		&\colhead{(7)}
		& \colhead{(8)}
		& \colhead{(9)}
		& \colhead{(10)}
	}
	\startdata
	IC~2038& {\it g} & $18.89\pm1.10$ & $1.01\pm0.09$ & $23.15\pm0.54$ & $13.38$ & $26.47\pm4.79$ & $22.39\pm0.13$ & $13.28$ & $52\%$&IC~2038&$0.41\pm0.08$&$0.62$&$4.41 \times 10^{8}$&$2.0 \times 10^{8}$\\
	& {\it r} & $19.47\pm0.39$ &  $0.99\pm0.02$  & $22.40\pm0.09$ & $12.57$ & $42.11\pm0.45$ & $22.88\pm0.10$  & $12.76$ & $45\%$&&&&\\
	\hline 
	IC~2039 & {\it g}  & $8.55\pm0.73$ & $1.61\pm0.11$  & $22.11\pm0.13$ & $14.06$ & $89.87\pm19$ & $24.36\pm0.16$ & $12.6$ & $79\%$&IC~2039&$0.71\pm0.20$&$1.53$&$1.36 \times 10^{9}$&$1.18 \times 10^{9}$\\
	& {\it r} & $10.18\pm0.24$ & $1.66\pm0.70$ & $21.89\pm0.61$ & $13.46$ & $132.26\pm19.50$ & $24.05\pm0.08$ & $11.45$ & $86\%$&&&&\\
	\enddata
	\tablecomments{(1) Filters. Best fit parameters of the light profile decomposition: (2) effective and scale radius;
		(3) S\'{e}rsic index, (4) effective and central surface brightness of the different components and (5) total magnitude of the individual component. (6) Total accreted mass fraction derived from our fit. (7) Averaged extinction-corrected $g-r$ color value and relative mass-to-light ratio in {\it r} band (8). Total stellar mass (9) and total accreted stellar mass fraction derived from the three-component fits (10) in {\it r} band.}
\end{splitdeluxetable*}

\begin{deluxetable*}{lccccccc}
	\tablecaption{GALFIT light profile decomposition parameters for {\it g} band}\label{tab:fit}
	\tablenum{6}
	\tablewidth{0pt}
	\tablehead{
		\colhead{Galaxy} 
		&\colhead{r$_{e}$} 
		& \colhead{n} 
		& \colhead{m$_{\text{Sérsic}}$} 
		&\colhead{r$_{0,1}$} 
		& \colhead{m$_{\text{exp,1}}$}
		& \colhead{r$_{0,2}$}
		& \colhead{m$_{\text{exp,2}}$}\\
		\colhead{} 
		&\colhead{[arcsec]} 
		&\colhead{}
		&\colhead{[mag]} 
		&\colhead{[arcsec]} 
		&\colhead{[mag]} 
		&\colhead{[arcsec]} 
		&\colhead{[mag]} \\
		\colhead{(1)} 
		&\colhead{(2)}
		&\colhead{(3)}
		&\colhead{(4)}
		&\colhead{(2)}
		&\colhead{(4)}
		&\colhead{(2)}
		&\colhead{(4)}
	}
	\startdata
	NGC~1533 & $5.76\pm0.01$ & $1.70\pm0.05$ & $12.31\pm0.06$ & $30.56\pm0.08$ & $11.31\pm0.04$ & $180.94\pm0.3$  & $11.65\pm0.05$ \\
	IC~2038& $39.61\pm2.15$ & $1.41\pm0.07$ & $14.02\pm0.05$ & \dots & \dots  & \dots & \dots \\
	IC~2039&$11.49\pm0.58$ & $1.98\pm0.07$ & $14.43\pm0.04$ & \dots & \dots  & \dots & \dots \\
	\enddata
	\tablecomments{(1) Galaxy name. Best fit parameters of the light profile decomposition: (2) effective and scale radius; (3) S\'{e}rsic index; (4) integrated magnitude of the different components.}
\end{deluxetable*}

\begin{figure*}
	\plotone{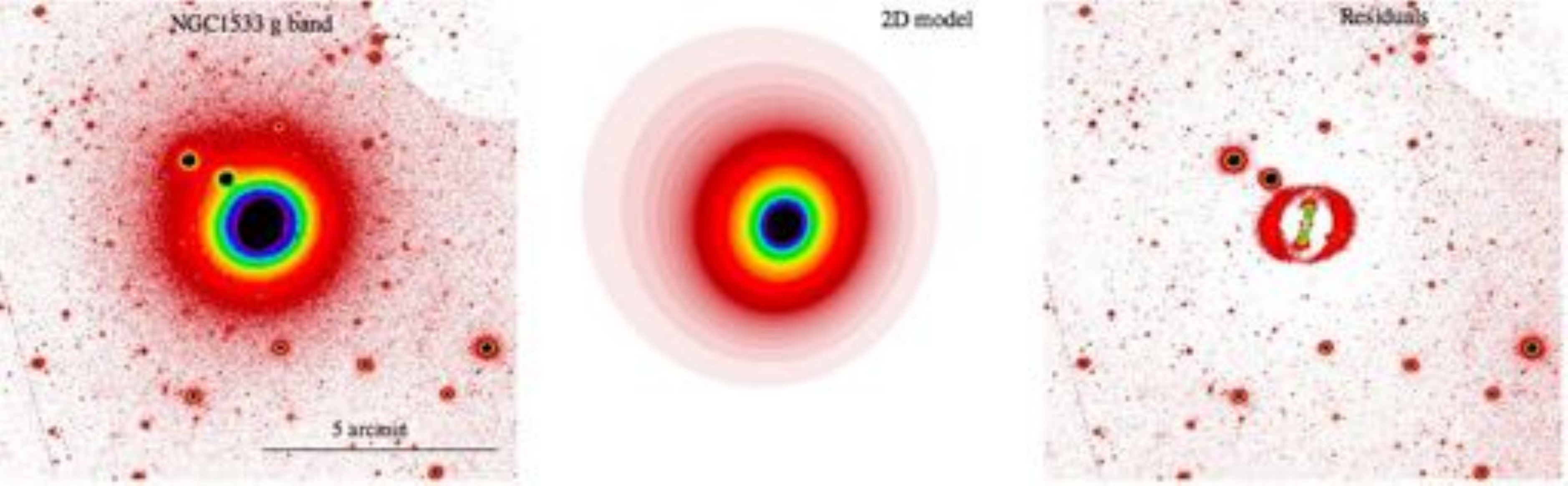}
	\plotone{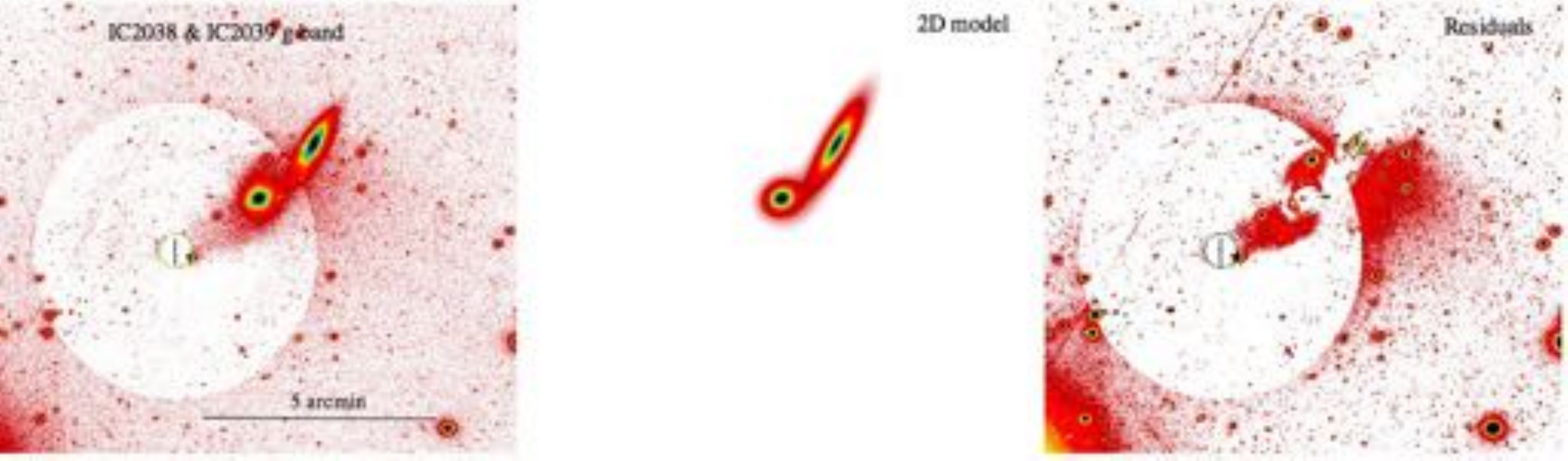}
	\caption{{\it Left panel}: {\it g} band image, {\it middle panel}: best fit model of the galaxies made with GALFIT, {\it right panel}: residual image obtained by subtracting the model from the {\it g} band image of NGC~1533 ({\it top panels}) and of IC~2038 and IC~2039 ({\it bottom panels}).\label{fig:fit2d}} 
\end{figure*}
\begin{figure}
	\gridline{\fig{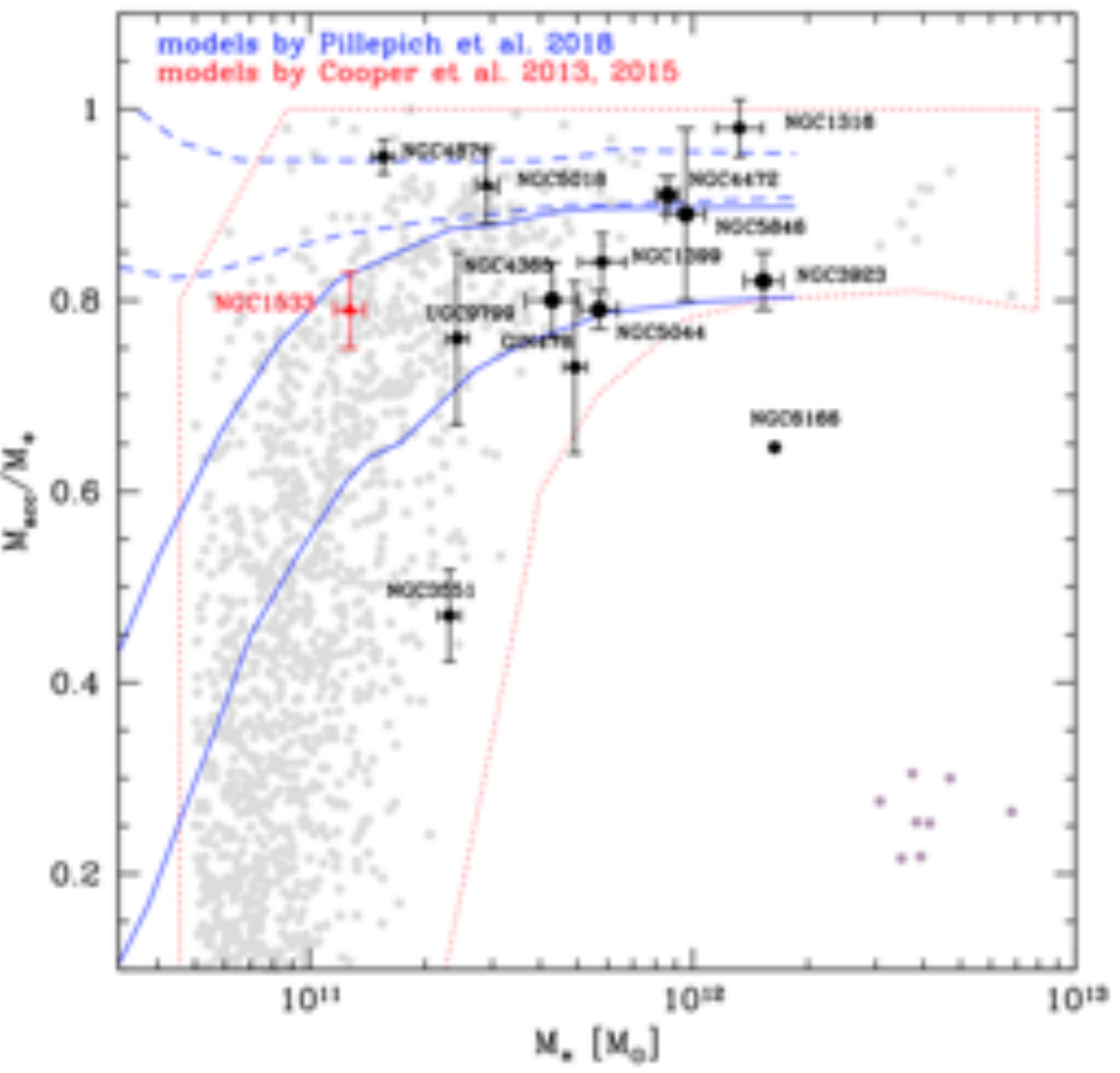}{0.46\textwidth}{}}
	\caption{Accreted mass fraction vs. total stellar mass for ETGs. The measurement for NGC~1533 is given as red triangle. Black circles correspond to other BCGs from the literature \citep{Seigar2007,Bender2015,Iodice2016,Iodice2017a,Spavone2017b,Spavone2018}. Red region indicates the predictions of cosmological galaxy formation simulations by \citet{Cooper2013,Cooper2015}. Blue continuous and dashed regions indicate the accreted mass fraction measured within $30$~kpc and outside $100$~kpc, respectively, in Illustris simulations by \citet{Pillepich2018} (see their Figure~$12$).Purple-gray points show the mass fraction associated with the streams from Table~$1$ in \citet{Cooper2015}.} \label{fig:halo}
\end{figure}

As done in \citet{Spavone2017b,Spavone2018}, in Figure~\ref{fig:halo} we compared the total accreted mass fraction estimated for NGC~1533 (see Table~\ref{tab:fit1d}),  with that predicted from theoretical models \citep{Cooper2013,Cooper2015,Pillepich2018} and with those derived from observations.

We found that the stellar mass fraction of the accreted component derived for NGC~1533 is fully consistent with other BCGs \citep[see][]{Seigar2007,Bender2015,Iodice2016,Spavone2017b,Spavone2018} and  with theoretical models from semi-analytic particle-tagging simulations \citep{Cooper2013,Cooper2015} and with Illustris cosmological hydrodynamical simulations \citep{Pillepich2018}. This further supports the mass assembly history traced above for NGC~1533.

Considering what we have just mentioned, and by comparing the morphology and light distribution of these set of simulations with those observed for NGC~1533 (see Sec.~\ref{sec:light}), we suggest a similar Aq-A model mass assembly history for the stellar halo in this galaxy.

\begin{deluxetable}{lccccc}
	\tablenum{7}
	\tablecaption{Basic properties of the dwarf galaxies around NGC~1533}\label{tab:dwarfs}
	\tablehead{
		\colhead{Galaxy}
		&\colhead{RA(J2000)}
		&\colhead{Decl.(J2000)}
		&\colhead{m$_{tot,g}$}
		&\colhead{$g-r$}\\
		\colhead{}
		&\colhead{}
		&\colhead{}
		&\colhead{[mag]}
		&\colhead{[mag]}\\
		\colhead{(1)}
		&\colhead{(2)}
		&\colhead{(3)}
		&\colhead{(4)}
		&\colhead{(5)}
	}
	\startdata
	LEDA~75089&$04^{h}11^{m}41^{s}.0$&$-56^{d}02^{m}55^{s}$&$17.4$&$0.7$\\
	LEDA~75054&$04^{h}08^{m}36^{s}.9$&$-55^{d}48^{m}15^{s}$&$18.4$&$1.1$\\
	LEDA~75094&$04^{h}12^{m}05^{s}.4$&$-55^{d}50^{m}54^{s}$&$18$&$1.2$\\
	LEDA~75095&$04^{h}12^{m}15^{s}.7$&$-55^{d}52^{m}07^{s}$&$16.9$&$0.5$\\
	\enddata
	\tablecomments{(1) Galaxies name; (2) and (3) J2000 coordinates from {\tt NED} based on the dwarf galaxies classification made by \citet{Ferguson1990}; (4) total magnitude in {\it g} band; (5) mean extinction corrected color value.}
\end{deluxetable}
\subsection{The complex merger history of NGC~1533 system: who is interacting with who?}

The analysis of the inner galaxy structure and the stellar envelope, given in the previous sections, suggests that NGC~1533 has a complex history of merging, probably with smaller-mass galaxies. The relics of these past interactions still remain in the inner regions of the galaxy and in the outskirts in the form of spiral-like structures and faint tails.

In Figure~\ref{fig:surbrig} is shown the HI emission detected by \citet{RyanWeber2003a}. To explain the HI gas ring-like formation, \citet{RyanWeber2003b} proposed the tidal destruction of a galaxy to form a merger remnant around NGC~1533.
Later, by looking at  the HI distribution and at the H$\alpha$ signatures, \citet{RyanWeber2004} also suggested that NGC~1533 could be the result of a minor, unequal-mass merger between a gas-rich low surface brightness galaxy and a disk high surface brightness galaxy. 
The merger transformed the disk galaxy into a barred lenticular galaxy, NGC~1533, and the outer HI disk of low-surface brightness galaxy into the giant HI ring around NGC~1533. Moreover \citet{RyanWeber2003a} and \citet{Bekki2005} would give reason to the surface brightness levels of the faint features in the galaxy outskirts.
According to this simulation, the lack of gas in the inner part of NGC~1533 confirms the hypothesis supported by UV emission and bluer structures that new stars are being generated in the stripped gas, rather than with material at the center of NGC~1533. 
On the other hand the HII confirmed signatures supports the hypothesis that new stars are being formed in the giant gas ring 
\citep{RyanWeber2003a,RyanWeber2003b,Werk2010}. Looking at the HI velocity dispersion and gradients, the information indicates that stars might have formed by cloud-to-cloud collisions in the HI envelope that is yet to stabilize \citep{RyanWeber2003b}.

Tacking into account the HI velocity one more possible explanation for the elliptical distributions could be that IC~2038, the small late-type galaxy, have interacted on a parabolic orbit with NGC~1533 in the past. During the interaction, gas and stars could have been stripped from the outskirts of IC~2038 along the orbit. This would also turn to be consistent with the presence of the arc-like structure detected on the West side of the group (see Figure~\ref{fig:surbrig}), which seems to follow a loop in the direction of NGC~1533.
The ATCA HI observations, which are sensitive to an area at least another $5\arcmin$ to West (compared with the plot~A in Figure~1 of \citet{RyanWeber2003b}), do not show HI emission in this region. It is quite possible that HI is present, but it is below the detection limit.

According to the stellar mass an estimate of the HI gas content of IC~2038 could be of the order of $\sim 10^9$~ $\mathcal{M}_{\odot}$ \citep{Denes2014,Denes2016}.
The interaction between NGC~1533 and IC~2038 was already proposed by \citet{Kilborn2005} and the present work put more constraints on it. Nevertheless IC~2038 does not show unambiguous signatures of interaction, although it seems it should be an obvious gas donor. Indeed IC~2039 is an ETG, which are typically gas-poor systems \citep{Young2018}. 
It is puzzling that  our deep imaging suggests that IC~2039 is significantly distorted (residual image from the 2-dimensional fit, bottom panel of Figure~\ref{fig:fit2d}), at odds IC~2038. It has clearly open arms generated by an on-going interaction, one of which is in the direction of IC~2038, the other in the opposite direction suggesting an interaction between them.

\section{Concluding summary}\label{sec:conc}

This work represent the first one of a series aimed to investigate the galaxy evolution in the Dorado group, a still un-virialized galaxy association \citep{Firth2006} whose bright galaxies show signatures of interaction. In this context, we have obtained new deep images of the field around one of the brightest group members, NGC~1533, as target of the VEGAS Survey.
Our deep surface photometry has been motivated by the fact that these galaxies appear embedded into a distorted HI envelope \citep{Kilborn2005} likely an evolving cloud within the group \citep{Werk2010,Rampazzo2017}. 
The present study suggests that NGC~1533 had a complex history made of several interactions with low-mass satellites that generated the star-forming spiral-like structure in the inner regions of the galaxy and are shaping the stellar envelope. 
In addition, the VST observations show that also the two less luminous members IC~2038 and IC~2039 are probably interacting each-other and, in the past, IC~2038 could have also interacted with NGC~1533, which stripped away gas and stars from its outskirts.

In conclusion, the three Dorado group members (NGC~1533, IC~2038 and IC~2039) may compose an interacting triplet, where the brightest galaxy NGC~1533 is ``red'' but not ``dead'' and with the ongoing mass assembly in the outskirts. Furthermore NGC~1533, IC~2038 and IC~2039 are having a quite eventful life likely representing one of the ``evolving roads'' of galaxies  in loose group. The study of both the  stellar and ionized gas kinematics will provide  important, still lacking, pieces in composing the puzzle of their co-evolution.


\acknowledgments 
The authors wish to thank the anonymous referee for his/her comments and suggestions that help us to improve the paper. The authors are grateful to L. Coccato for him helpful contribution in developing the PSF deconvolution task. AC, EI and MS acknowledge financial support from the VST project (P.I. P. Schipani). RR and EVH acknowledges funding from the PRIN-INAF eSKApe HI 2017 program 1.05.01.88.04. AC thanks E.M. Corsini, E. Sissa, M. Berton and E. Congiu for useful suggestions. This research has made use of the NASA/IPAC Extragalactic Database ({\tt NED}), which is operated by the Jet Propulsion Laboratory, California Institute of Technology, under contract with the National Aeronautics and Space Administration. We acknowledge the usage of the {\tt HyperLeda} database  (http://leda.univ-lyon1.fr). IRAF is distributed by the National Optical Astronomy Observatories, which are operated by the Association of Universities for Research in Astronomy, Inc., under cooperative agreement with the National Science Foundation. The VST project is a joint venture between ESO and the National Institute for Astrophysics (INAF) in Naples, Italy. 

%

\vspace{5mm}
\facilities{ESO-VST, OmegaCam@ESO-VST}


\software{IRAF, MINUIT, GALFIT, HyperLeda}

\appendix

\section{Isophotal analysis}\label{apx:A}

\subsection*{NGC~1533}
The analysis of the isophote fit allows us to make the following considerations (see Figure~\ref{fig:geompar}): {\it (i)} the bulge is located in the region $0\farcs79<R<6\farcs95$. It shows a variation in the position angle, $107^{\circ}-134^{\circ}<$P.A.$<152^{\circ}$, and in ellipticity, $0.033<\epsilon<0.102$ and it has roughly zero flat {\it a4} and {\it b4} profiles. {\it (ii)} The bar plus lens component is located in the range $6\farcs95<R<46\farcs7$. It has a variation in the position angle, $152^{\circ}<$P.A.$<166^{\circ}$ for $6\farcs95<R<29\arcsec$ and $102^{\circ}<$P.A.$<166^{\circ}$ for $29\arcsec<R<46\farcs7$. The ellipticity increases, $0.102<\epsilon<0.369$, for $6\farcs95<R<29\arcsec$ then it decreases, $0.055<\epsilon<0.365$, for $29\arcsec<R<46\farcs7$. {\it (iii)} The outer ring-lens structure is located in the region $46\farcs7<R<82\farcs58$; it has a variation in the position angle, $108^{\circ}<$P.A.$<122^{\circ}$, and in ellipticity, $0.055<\epsilon<0.085$ and it has maximum values $\sim0.005$ and $\sim0.01$ of {\it a4} and {\it b4}, respectively. {\it (iv)} The disk plus stellar envelope component is located out of $R>82\farcs58$. 

Based on the isophotal analysis, the azimuthally averaged surface brightness profile is best reproduced by a multi-component model, the profile is fitted with three components: two S\'{e}rsic and an outer exponential function. These three elements define the different light components dominating the galaxy light distribution at different scales (Figure \ref{fig:fit1d} and Table \ref{tab:fit1d}).

\subsection*{IC~2038}
The geometrical parameters of IC~2038 confirm that the galaxy is a late-type (Figure \ref{fig:geompar}). The position angle is approximately constant around $\sim145^{\circ} - 155^{\circ}$, while the ellipticity has a rapid increase, from $0.2$ to $0.7$, in the central regions, between $0\farcs79$ and $4\farcs2$, and then it sets in a range of $\sim 0.6 - 0.8$ outside $12\arcsec$. This behavior highlights a disky nature that is confirmed by the positive values, in all galaxy extension, of {\it b4} profile. The large error bars of these two last parameters are compliant with the presence of dust .

The azimuthally averaged surface brightness profile is best reproduced by a multicomponent model: the profile is fitted with a S\'{e}rsic and an outer exponential function, as shown in Figure \ref{fig:fit1d} and Table \ref{tab:fit1dIC}.

\subsection*{IC~2039}
\begin{figure}[h]
	\gridline{\fig{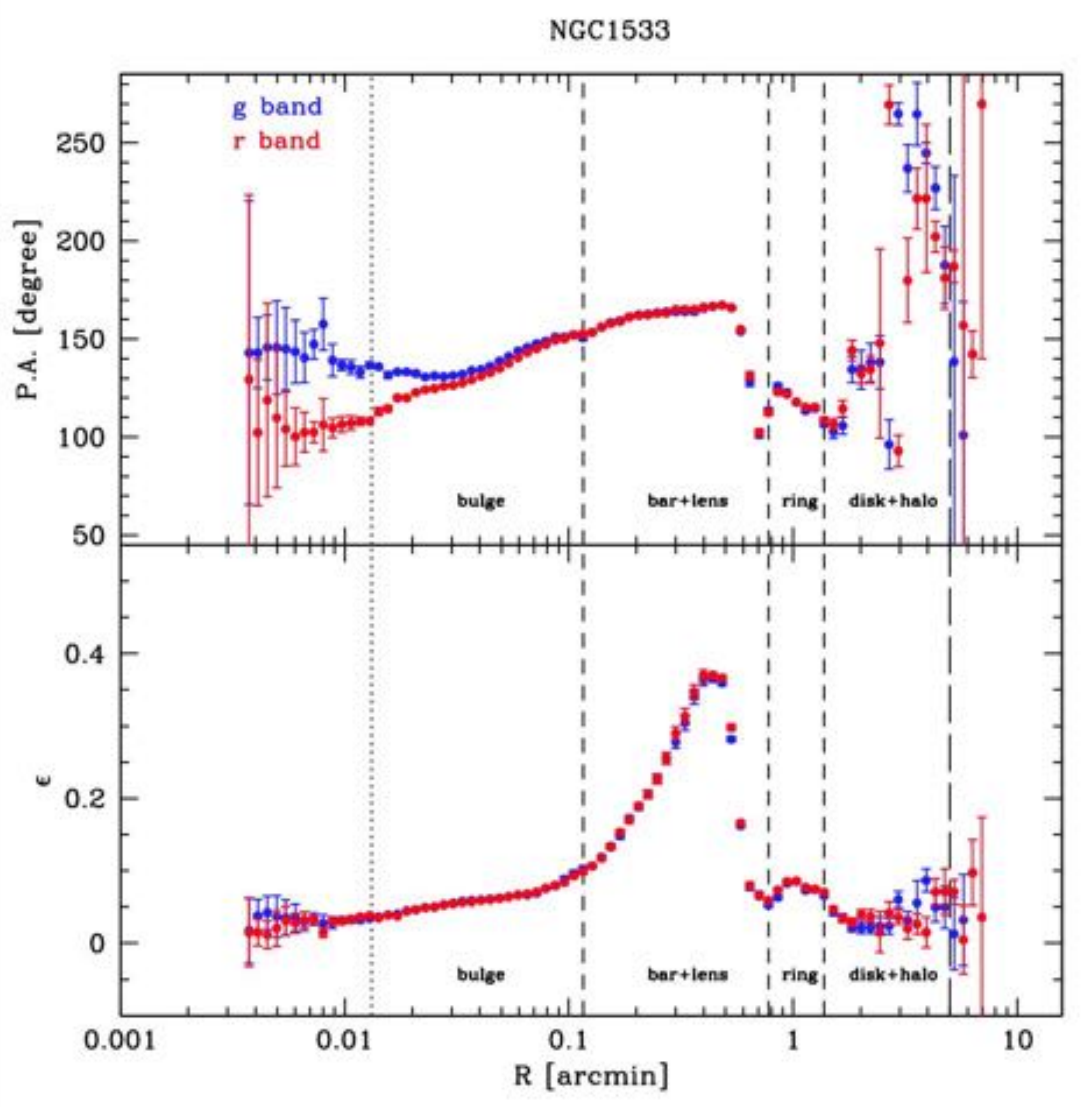}{0.35\textwidth}{}
		\fig{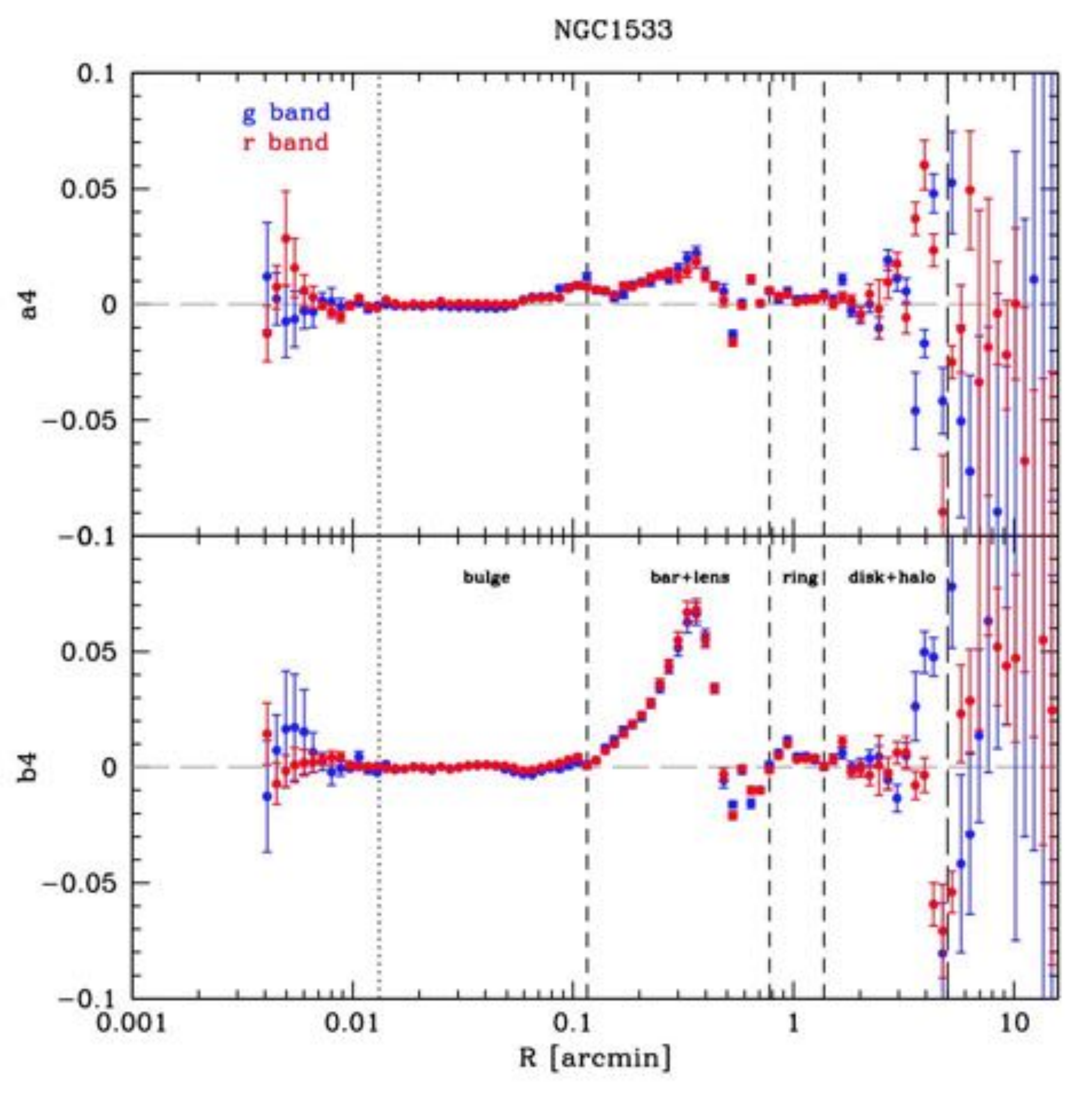}{0.35\textwidth}{}
	}
	\gridline{\fig{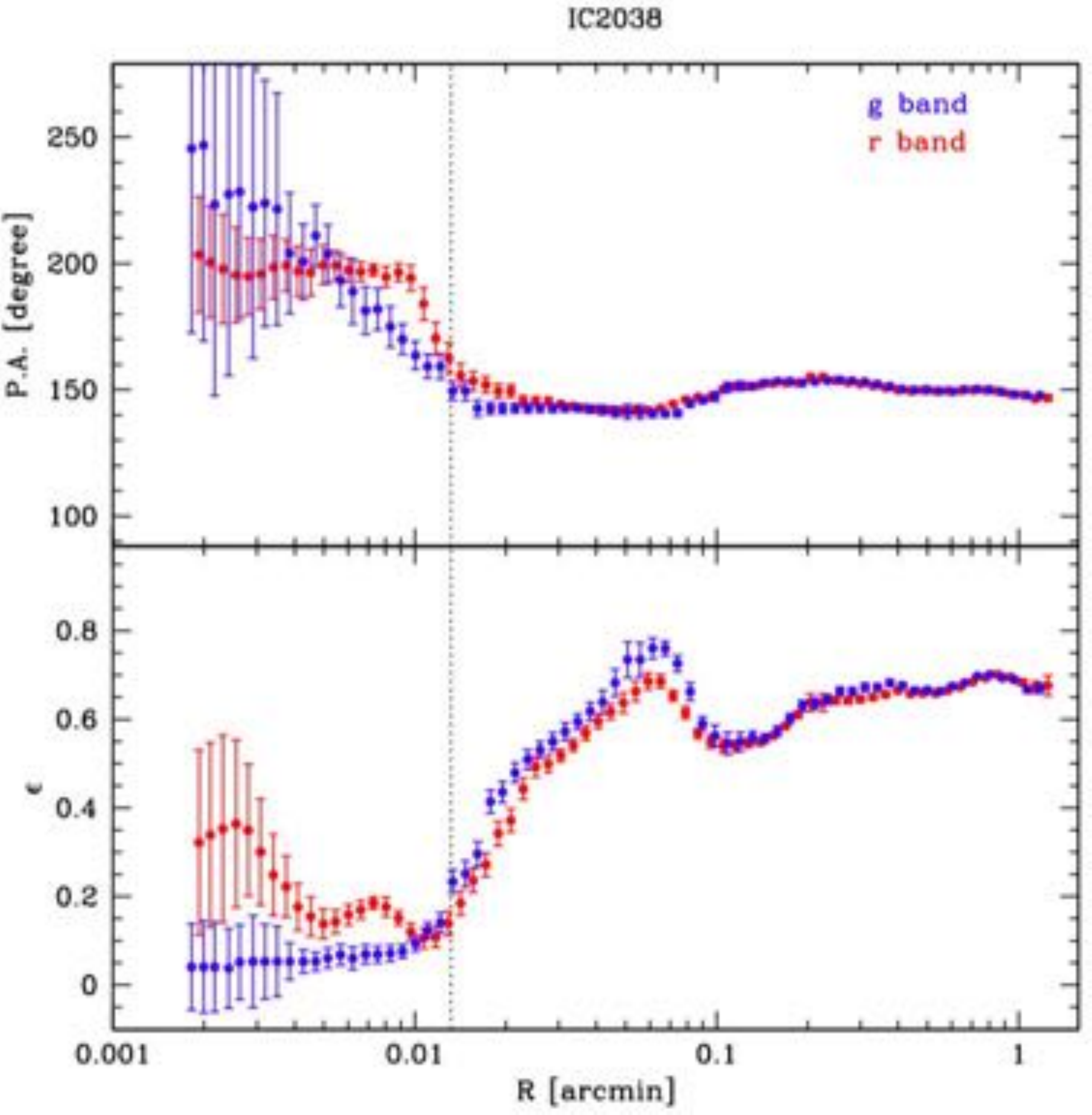}{0.34\textwidth}{}
		\fig{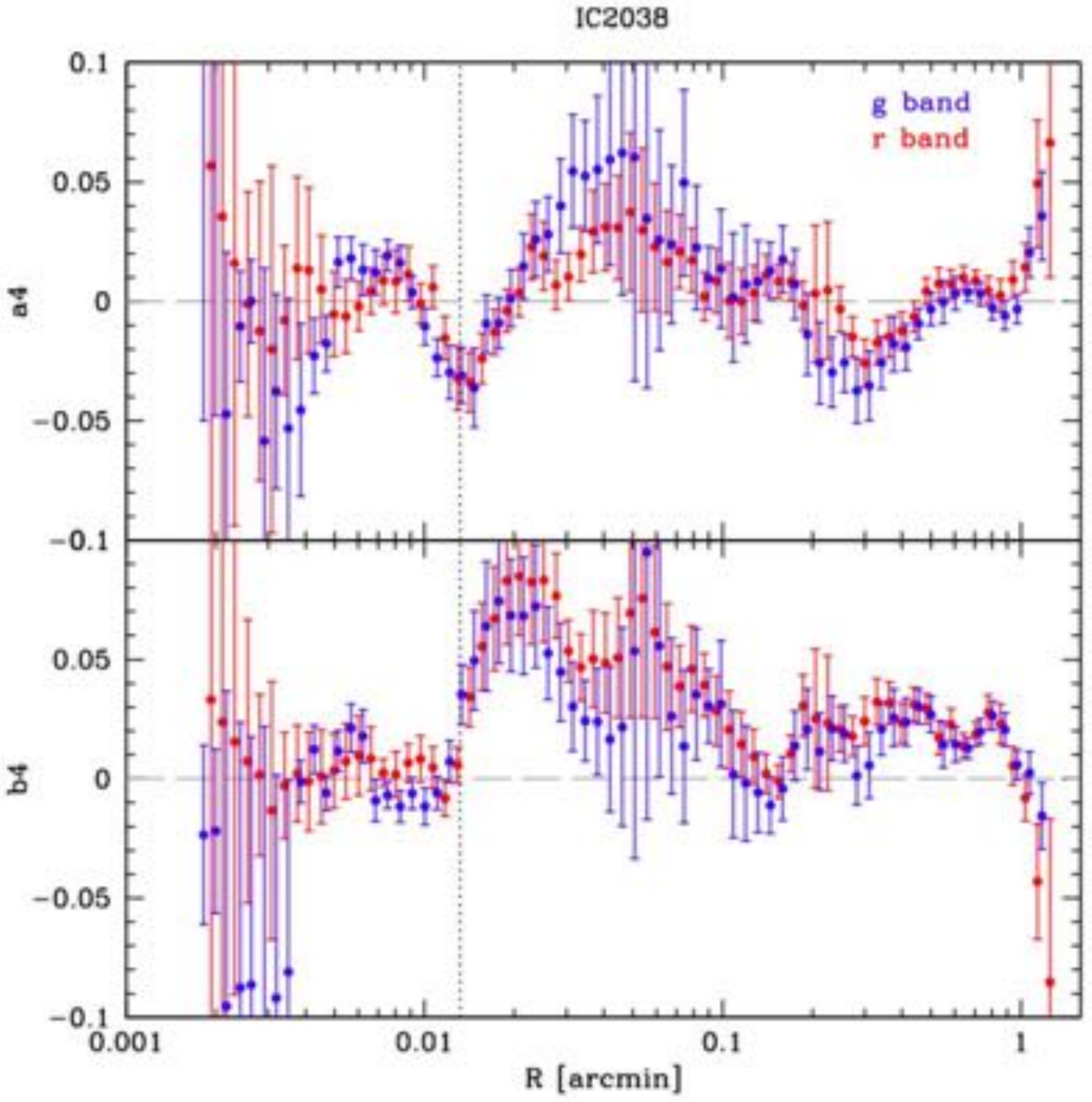}{0.34\textwidth}{}
	}
	\gridline{\fig{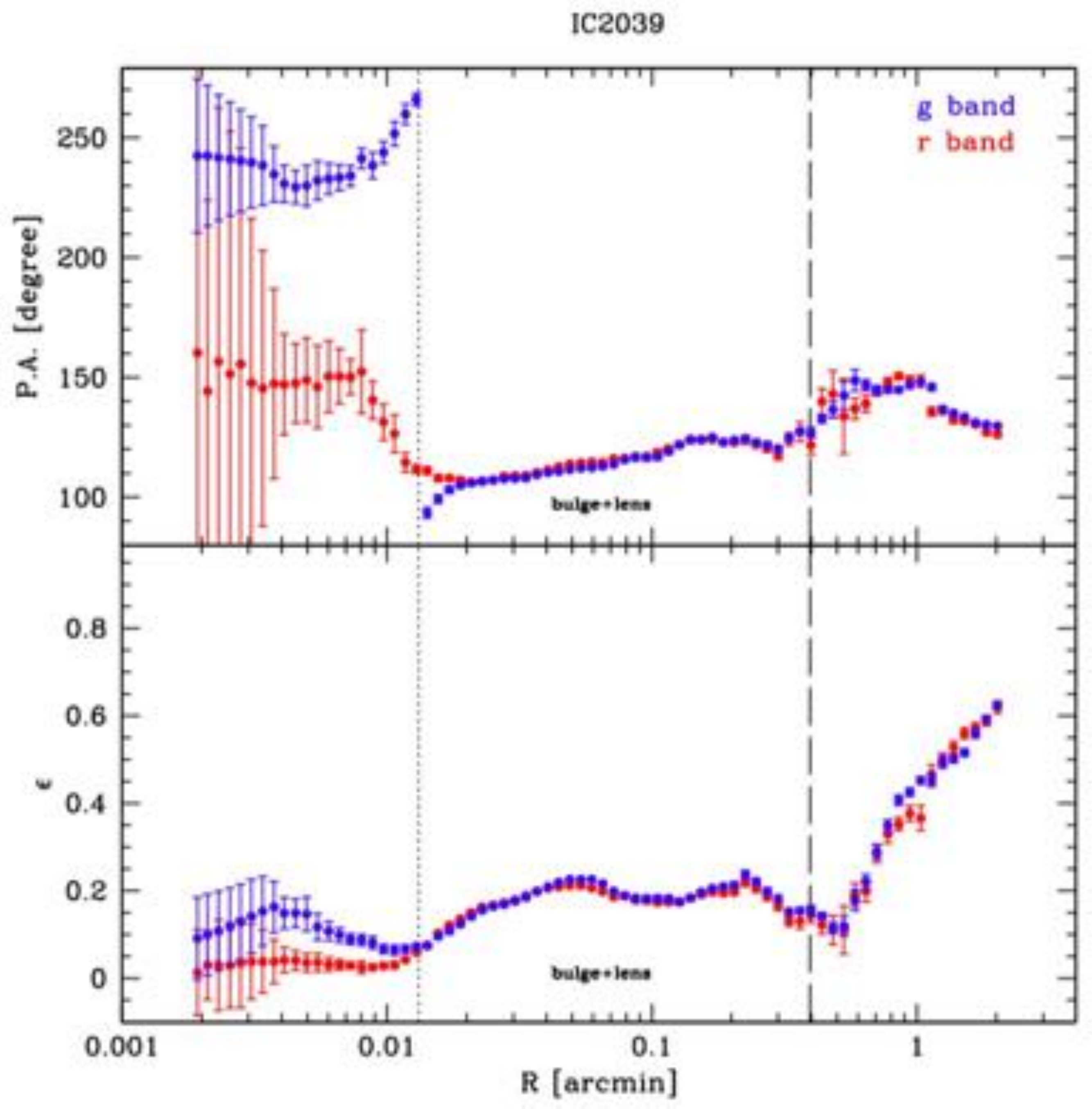}{0.34\textwidth}{}
		\fig{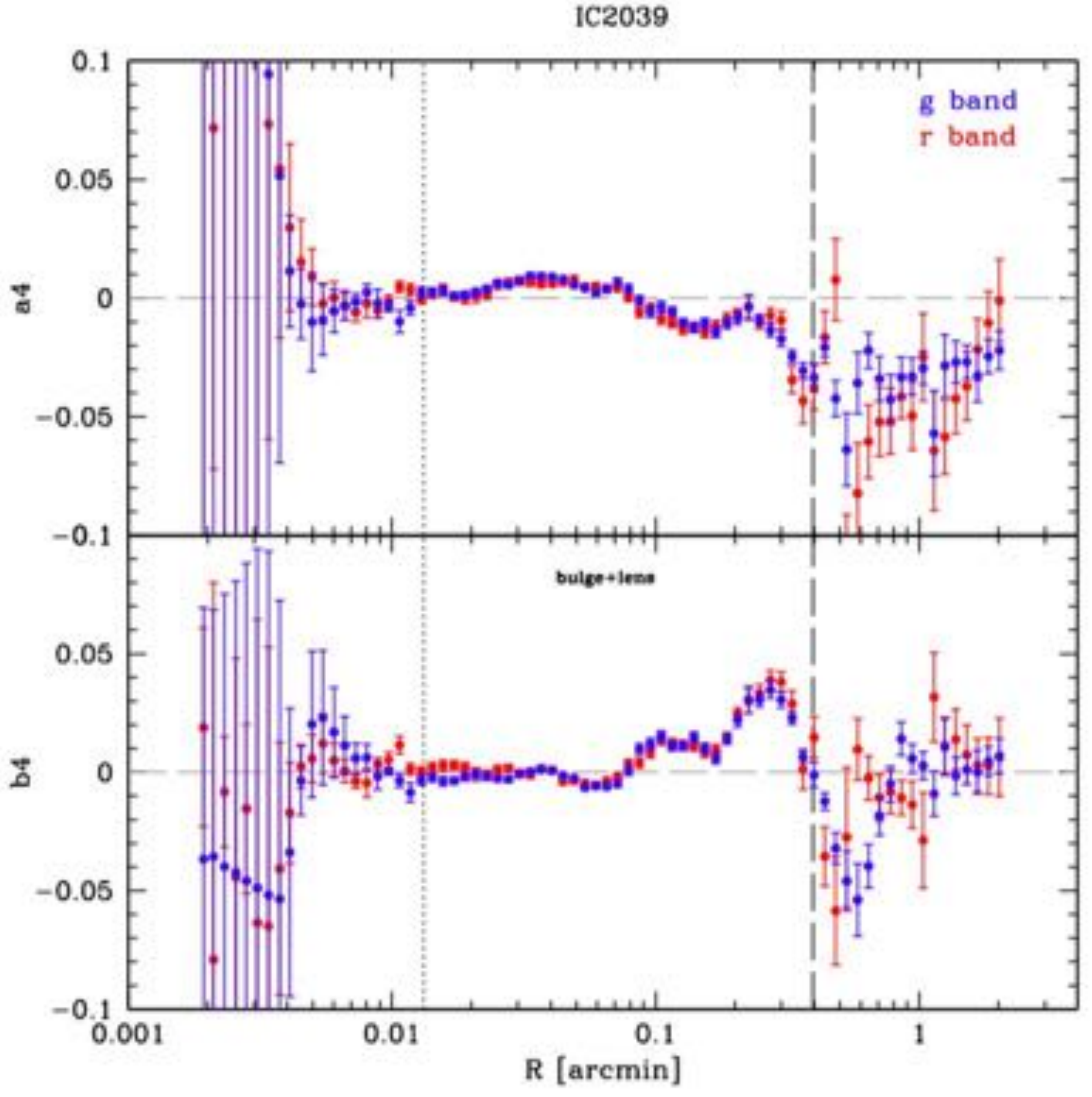}{0.34\textwidth}{}
	}
	\caption{Position angle ({\it top left panel}), ellipticity ({\it bottom left panel}), $a4$ ({\it top right panel}) and $b4$ ({\it bottom right panel}) profiles plotted against the isophote semi-major axis. These plots are obtained with \textsc{ellipse} from the {\it g} band image, blue dots, and {\it r} band image, red dots.\label{fig:geompar}}
\end{figure}

\begin{figure}
	\gridline{\fig{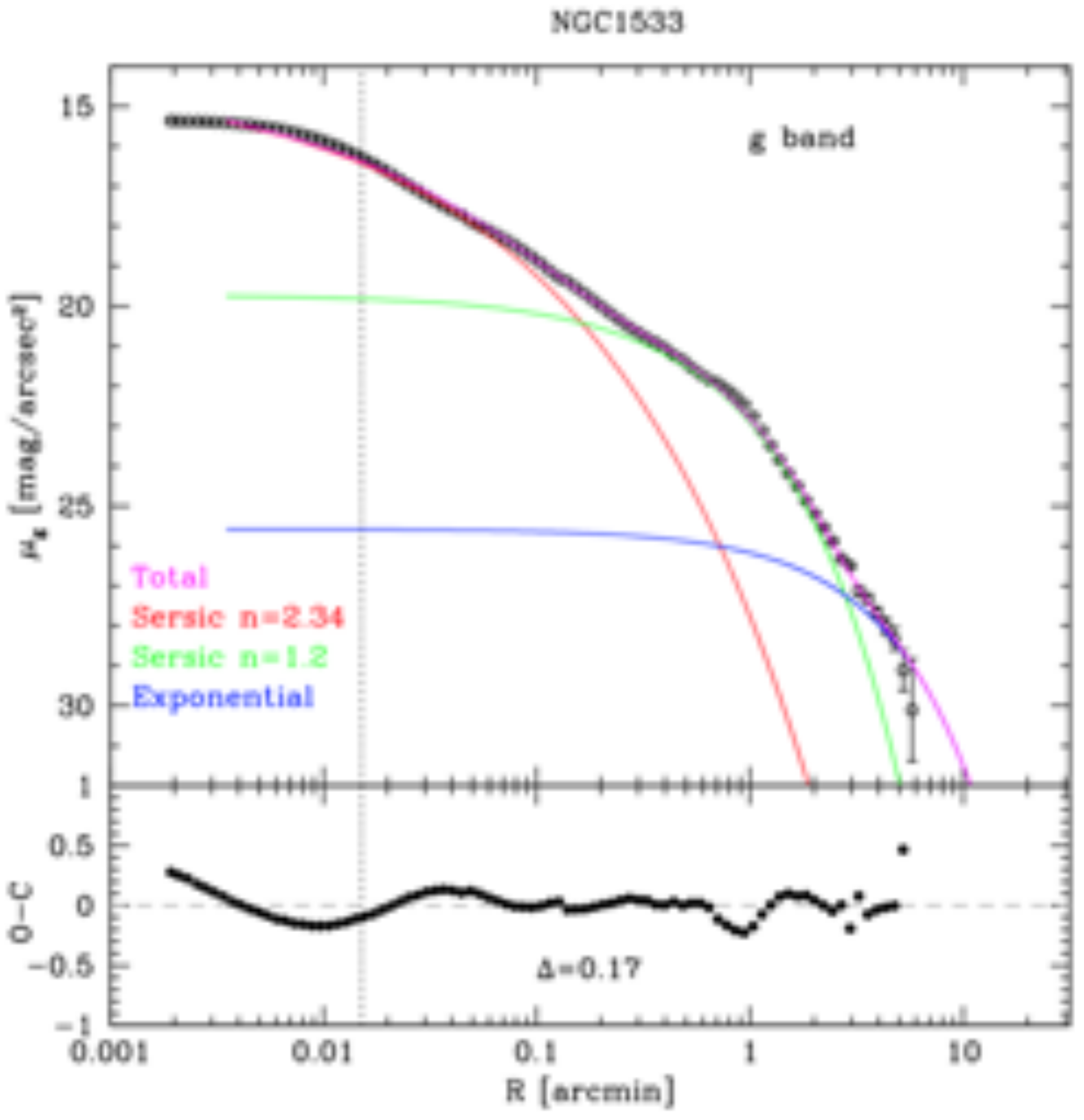}{0.34\textwidth}{}
		\fig{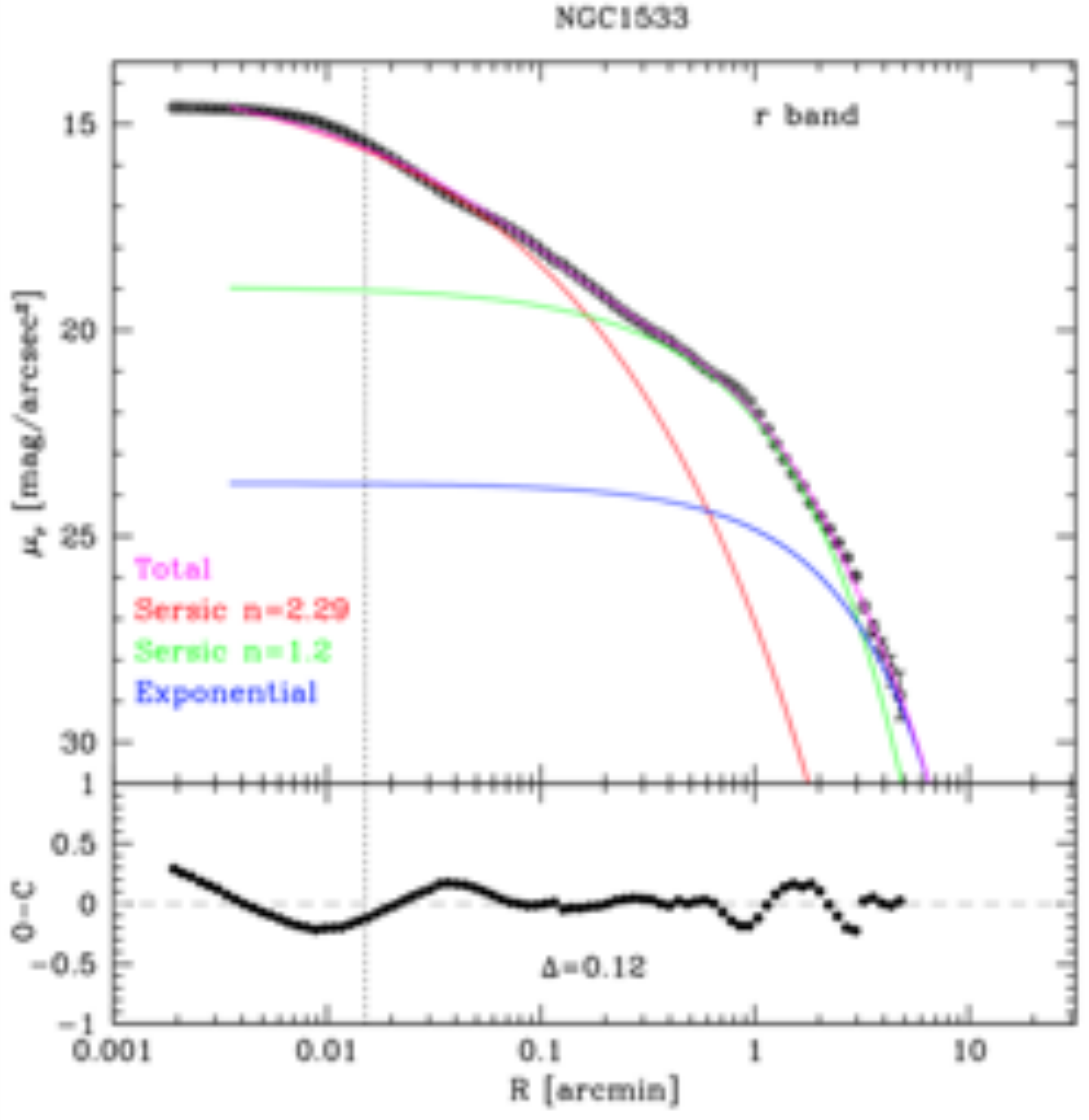}{0.35\textwidth}{}
	}
	\gridline{\fig{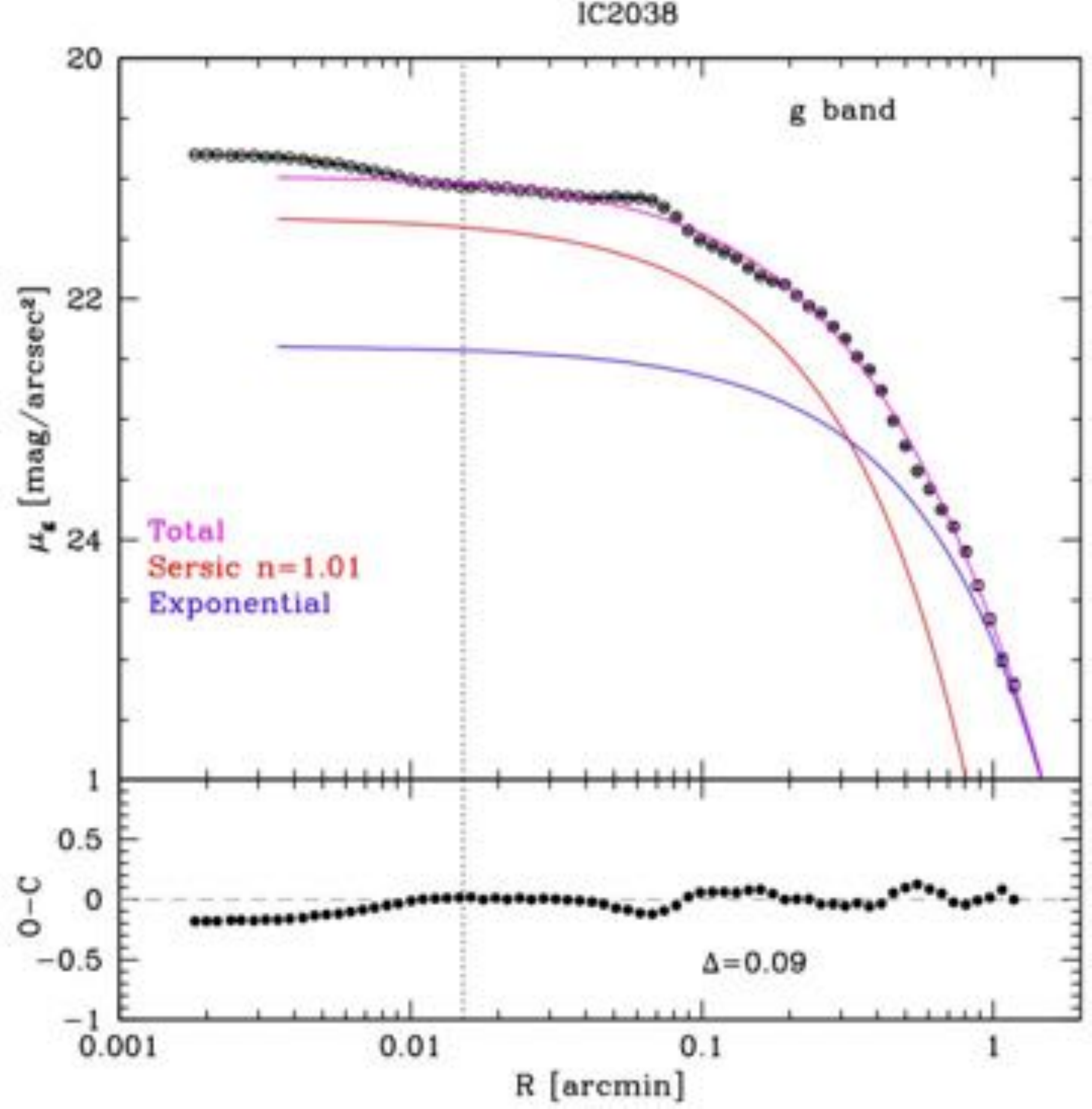}{0.34\textwidth}{}
		\fig{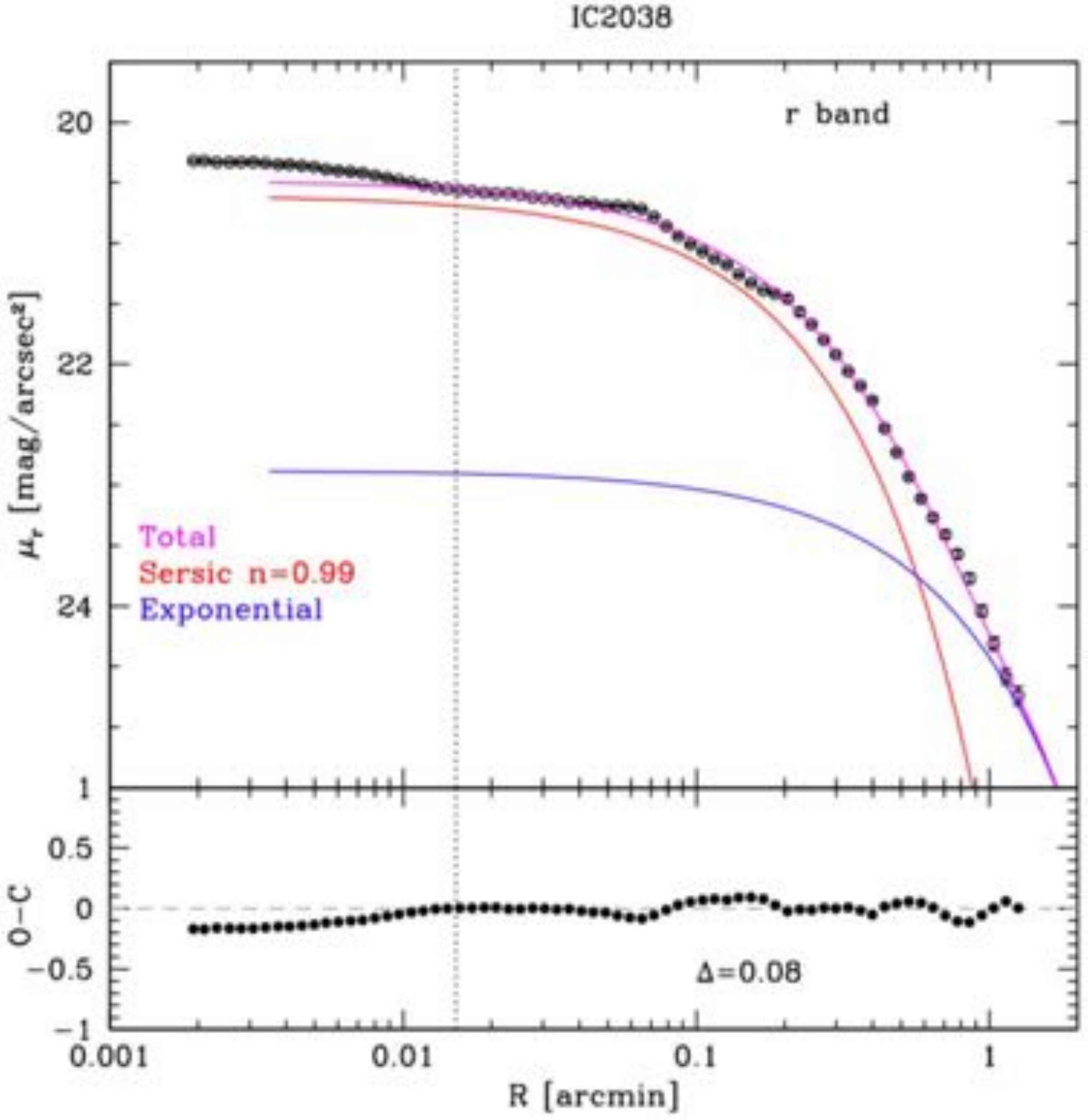}{0.34\textwidth}{}
	}
	\gridline{\fig{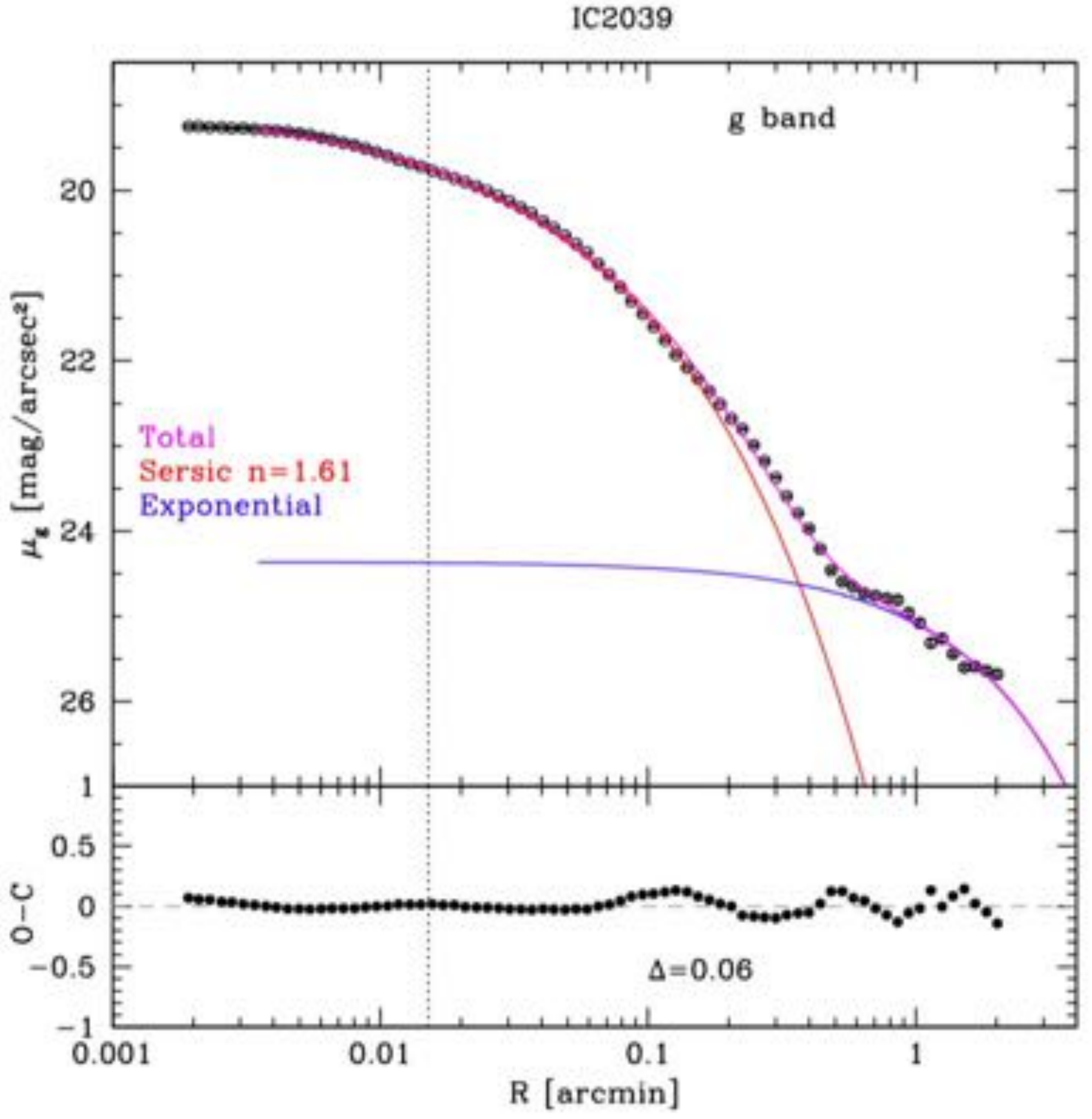}{0.34\textwidth}{}
		\fig{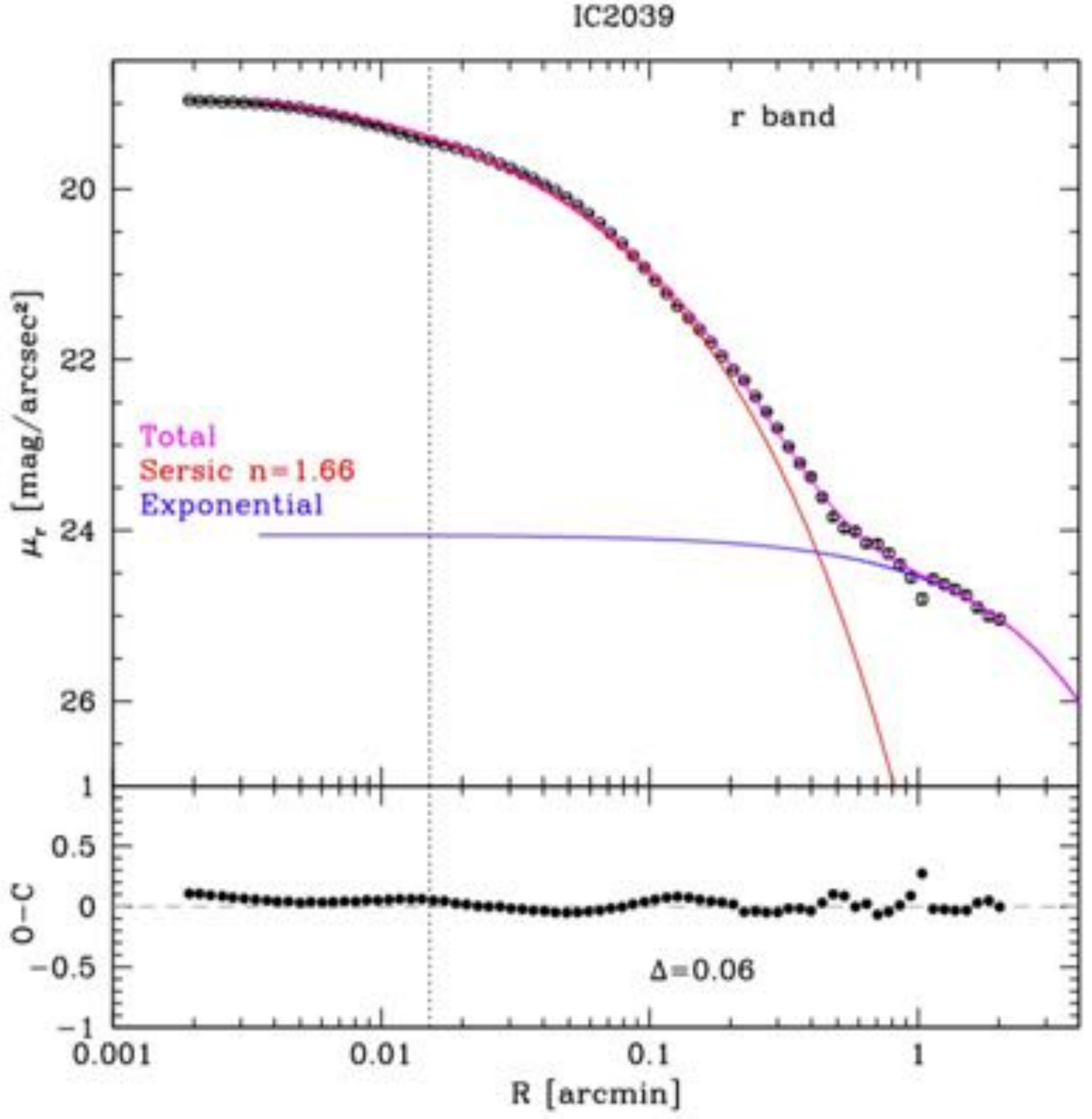}{0.34\textwidth}{}
	}
	\caption{The azimuthally averaged (and PSF deconvolved for NGC~1533) surface brightness profiles in the {\it g} band ({\it left panels}) and in {\it r} band ({\it right panels}) and the residuals between the observed surface brightness and the best fit, the gray dashed line indicates the zero level. {\it Top panels} refer to NGC~1533, {\it middle panels} to IC~2038 and {\it bottom panels} to IC~2039. Red and green solid lines represent S\'{e}rsic laws, blue solid line exponential function and magenta solid line is the resulting best fit interpolation. The vertical dotted line delimits the regions affected by seeing, which is excluded from the fit.\label{fig:fit1d}}
\end{figure}

Looking at Figure~\ref{fig:geompar} (bottom panels), IC~2039 presents an ellipticity lower than $\sim 0.25$ up to $42\arcsec$ and a constant increase to $\sim 0.8$ at about $2\arcmin$. On the contrary the position angle has a first roughly linear increasing trend, $\sim100^{\circ} - 125^{\circ}$, inside $18\arcsec$ and then it scatters from $100 ^{\circ}$ to $150 ^{\circ}$. 
The complexity of the structure is consistent with the {\it a4} and {\it b4} profiles. In particular $b4$ shows a inner flat component, supposedly the bulge, a positive trend between $4\farcs8 \leq R \leq 24\arcsec$, probably due to a disky component. 
We reproduced the azimuthally averaged surface brightness profile  using a S\'{e}rsic and an exponential function. These two elements should define the bulge light component and an outer disk feature (Figure \ref{fig:fit1d} and Table \ref{tab:fit1dIC}).

\end{document}